\documentclass[letterpaper,twocolumn,10pt]{article}
\usepackage{usenix2019_v3}
\makeatletter
\newcommand{\printfnsymbol}[1]{%
  \textsuperscript{\@fnsymbol{#1}}%
}
\makeatother
\usepackage{booktabs}       %
\usepackage{comment}        %
\usepackage[normalem]{ulem} %
\usepackage{subcaption}     %
\usepackage{url}            %
\usepackage{xspace}         %
\usepackage{tikz}           %
\usepackage{mathtools}
\usepackage{titling}      %
\usepackage{listings}        %
\usepackage{dsfont}
\usepackage{amsmath}
\usepackage[title]{appendix}
\usepackage{titling}      %
\usepackage[compact]{titlesec}
\usepackage{etoolbox}

\definecolor{codegreen}{rgb}{0,0.6,0}
\definecolor{codeblue}{rgb}{0,0.5,1.0}
\definecolor{codegray}{rgb}{0.5,0.5,0.5}
\definecolor{codepurple}{rgb}{0.58,0,0.82}
\definecolor{backcolour}{rgb}{0.95,0.95,0.92}

\lstdefinestyle{mystyle}{
    commentstyle=\color{codeblue},
    keywordstyle=\color{magenta},
    numberstyle=\tiny\color{codegray},
    stringstyle=\color{codepurple},
    basicstyle=\ttfamily\footnotesize,
    breakatwhitespace=false,
    breaklines=true,
    captionpos=b,
    keepspaces=true,
    numbers=left,
    numbersep=5pt,
    showspaces=false,
    showstringspaces=false,
    showtabs=false,
    tabsize=2,
}

\usepackage{subcaption}
\usepackage[shortcuts]{extdash} %

\usepackage{pifont}

\usepackage{algorithm}
\usepackage[noend]{algpseudocode}
\makeatletter
\renewcommand{\ALG@beginalgorithmic}{\footnotesize}
\makeatother

\usepackage{tabu}
\usepackage{makecell}
\usepackage{colortbl}
\usepackage{multirow}
\definecolor{TableRowColor}{rgb}{0.88,1,1}
\newcolumntype{?}{!{\vrule width 1pt}}

\newcommand{\infaas}{$\mathtt{INFaaS}$\xspace}
\newcommand{\binfaas}{\textbf{\texttt{INFaaS}}\xspace}
\newcommand{\modvar}{model\-/variant\xspace}
\newcommand{\modvars}{model\-/variants\xspace}
\newcommand{\Modvar}{Model\-/variant\xspace}
\newcommand{\Modvars}{Model\-/variants\xspace}

\newcommand{\staticgpu}{$\mathtt{Clipper}^{+}_{\mathtt{GPU}}$\xspace}
\newcommand{\staticcpu}{$\mathtt{Clipper}^{+}_{\mathtt{CPU}}$\xspace}
\newcommand{\indivscale}{$\mathtt{SM}^+$\xspace}
\newcommand{\indivgpu}{$\mathtt{SM}^{+}_{\mathtt{GPU}}$\xspace}
\newcommand{\indivcpu}{$\mathtt{SM}^{+}_{\mathtt{CPU}}$\xspace}
\newcommand{\static}{$\mathtt{Clipper}^+$\xspace}
\newcommand{\vanillasagemaker}{$\mathtt{SageMaker}$\xspace}
\newcommand{\vanillaclipper}{$\mathtt{Clipper}$\xspace}

\usepackage{caption}
\usepackage{enumitem}
\setitemize{nosep, leftmargin=1em}
\setenumerate{nosep, leftmargin=1em}

\usepackage{amssymb}

\begin{document}

\setlength{\droptitle}{-0.55in}

\title{\Large\bf \binfaas: A \emph{Model-less} and \emph{Managed} Inference Serving System}

\author{
  \normalsize Francisco Romero\thanks{Equal contribution} , Qian Li\printfnsymbol{1}, Neeraja J. Yadwadkar, Christos Kozyrakis \\
  \normalsize faromero@stanford.edu, qianli@cs.stanford.edu, neeraja@cs.stanford.edu, kozyraki@stanford.edu \\
  \normalsize Stanford University
}
\date{}
\maketitle

\begin{abstract}

Despite existing work in machine learning inference serving, \emph{ease-of-use} and \emph{cost efficiency} remain challenges at large scales. 
Developers must manually search through thousands of \emph{\modvars} -- versions of already-trained models that differ in hardware, resource footprints, latencies, costs, and accuracies -- to meet the diverse application requirements.
Since requirements, query load, and applications themselves evolve over time, these decisions need to be made dynamically for \emph{each} inference query to avoid excessive costs through naive autoscaling.
To avoid navigating through the large and complex trade-off space of \modvars, developers often fix a variant across queries, and replicate it when load increases. 
However, given the diversity across variants and hardware platforms in the cloud, %
a lack of understanding of the trade-off space can incur significant costs to developers.

This paper introduces \infaas, a \emph{managed} and \emph{model-less} system for distributed inference serving, where developers simply specify the performance and accuracy requirements for their applications without needing to specify a specific model-variant for each query. 
\infaas generates \modvars, and efficiently navigates the large trade-off space of \modvars on behalf of developers to meet application-specific objectives: (a) for each query, it selects a model, hardware architecture, and model optimizations, %
(b) it combines VM-level horizontal autoscaling with model-level autoscaling, where multiple, different model-variants are used to serve queries within each machine.
By leveraging diverse variants and sharing hardware resources across models, \infaas achieves 1.3$\times$ higher throughput, violates latency objectives 1.6$\times$ less often, and saves up to 21.6$\times$ in cost (8.5$\times$ on average) compared to state-of-the-art inference serving systems on AWS EC2. 

\end{abstract}

\section{Introduction}
\label{sec:introduction}
The number of applications relying on inference from Machine Learning (ML) models, such as video analytics~\cite{videostorm}, is already large~\cite{pywren, Scanner, chameleon, nlp_1, nlp_2} and expected to keep growing. 
Facebook, for instance, serves tens-of-trillions of inference queries per day~\cite{AppliedML_FB, arch-rmc}. %
Consequently, distributed inference dominates ML production costs: on AWS, inference accounts for over 90\% of ML infrastructure cost~\cite{inf_cost}. 

Inference serving is user-facing. 
Typically, an ML lifecycle has two distinct phases – training and inference. 
The training phase is usually characterized by long-running hyperparameter searches, dedicated hardware resource usage, and no completion deadlines. 
In the inference phase, trained models can be queried by various end-user applications. 
Being user-facing, inference serving requires cost-effective systems that render predictions with latency constraints while handling unpredictable and bursty request arrivals.

\begin{table}[t!]
  \centering 
  \small
\begin{tabular}{lccc}
\hline
Application        & Accuracy & Latency & Cost \\ \hline
\rowcolor[HTML]{CBCEFB} 
Social Media       & High      & Medium     & Low  \\
Visual Guidance    & High     & Low     & High \\
\rowcolor[HTML]{CBCEFB} 
Intruder Detection & Low      & Low     & Low  \\ \hline
\end{tabular}
    \caption{Applications querying a face recognition model with diverse requirements. }
    \label{tab:facemodel_example}
\end{table}

Inference serving systems face a number of challenges~\cite{hotos_anonymized, mlperf} due to the following factors.  \newline%
\noindent\textbf{(a) Diverse application requirements:} Applications issue queries that differ in latency, cost, accuracy, and even privacy~\cite{shredder} requirements~\cite{focus,OneSizeAll,mlperf}.
Table~\ref{tab:facemodel_example} shows the same face recognition model queried by multiple applications with different requirements. 
Some applications, such as intruder detection, require inference in realtime but can tolerate lower accuracy. 
Others, such as tagging faces on social media, may prefer accuracy over latency. \newline
\noindent\textbf{(b) Heterogeneous execution environments:} Leveraging heterogeneous hardware resources (e.g., different generations of CPUs, GPUs, and accelerators like TPU~\cite{TPU_ISCA} or AWS Inferentia~\cite{Inferentia}) helps meet the diverse needs of applications and the dynamic changes in the workload; however, it is non-trivial to manage and scale heterogeneous resources~\cite{arch-rmc}. \newline
\noindent\textbf{(c) Diverse \modvars:} Methods, such as layer fusion or quantization~\cite{TVM, tensorRT, aws_neuron}, produce versions of the same model, \emph{\modvars}, that may differ in inference latency, memory footprint, and accuracy. 

Together, these factors create a large search space. 
For instance, from 21 already-trained image classification models, we generated 166 \modvars, by (i) applying model graph optimizers, such as TensorRT~\cite{tensorRT}, (ii) optimizing for different batch sizes, and (iii) changing underlying hardware resources, such as CPUs, GPUs, and Inferentia.  
These variants vary across many dimensions: the accuracies range from 56.6\% to 82.5\% (1.46$\times$), the model loading latencies range from 590ms to 11s (18.7$\times$), and the inference latencies for a single query range from 1.5ms to 5.7s (3,700$\times$). 
Their computational requirements range from 0.48 to 24 GFLOPS (50$\times$)~\cite{mlperf}, and the cost of hardware these variants incur~\cite{EC2_price} ranges from \$0.096/hr for 2 vCPUs to \$3.06/hr for a V100 GPU (32$\times$). 
As new inference accelerators are introduced and new optimization techniques emerge, the number of model-variants will only grow. 

This large search space makes it hard for developers to manually match the requirements of each inference query to decisions about selecting the right model and model optimizations, suitable hardware platforms, and auto-scaling configurations.
The decision complexity is further exacerbated when the load varies, applications evolve, and the availability of hardware resources (GPUs, ASICs) changes. %
Unlike long-running batch data analytics or ML training jobs~\cite{Tiresias,themis,cherrypick,ernest,hydra} that can be right-sized during or across subsequent executions, the dynamic nature of distributed inference serving makes it infeasible to select \modvars statically. %

Our key insight is that the large diversity of \modvars is not a nuisance but an opportunity: it can allow us to meet the diverse and varying performance, cost, and accuracy requirements of applications, in the face of varying load and hardware resource availability, if only we are able to select and deploy the right model-variant effectively for \emph{each} query. 
However, given the complexity of this search space, existing systems, including Clipper~\cite{Clipper}, TensorFlow Serving~\cite{TensorFlowServing}, AWS SageMaker~\cite{SageMaker}, and others~\cite{TRT_Server,CloudML,Azure_ML,mark}, ignore the opportunity. %
These systems tightly couple a model to a hardware resource and use statically-defined resource management policies. %
These existing systems require developers to select \modvars, batch sizes, instance/hardware types, and autoscaling configurations, for meeting requirements of each query.  
If these decisions are made without understanding the trade-offs offered by the variants, the impact could be significant (note the wide cost, performance, and accuracy ranges spanned by the variants). %
We argue that in addition to traditional autoscaling, distributed inference serving systems should navigate this search space of \modvars on behalf of developers, and automatically manage \modvars and heterogeneous resources to meet the requirements of inference-driven applications. 
Surprisingly, as we also noted in~\cite{hotos_anonymized}, no existing inference serving system does that. %

To this end, we built \infaas (\underline{INF}erence-\underline{a}s-\underline{a}-\underline{S}ervice), a \emph{model-less} and \emph{managed} system for distributed inference serving. 
\infaas introduces a \emph{model-less} interface where after registering trained models, developers do not need to generate, select, or manage the \modvars for their applications. %
Instead, for each inference query they specify only the high-level performance, cost, or accuracy requirements. %
\infaas generates \modvars of the registered models, and navigates the large space to select a \modvar and automatically switch between differently optimized model-variants to best meet the goals of each query. 
\infaas is \emph{managed}, because it automates resource provisioning for \modvars and schedules queries across a heterogeneous cluster. %

To realize this, \infaas generates \modvars and their performance-cost profiles on different hardware platforms. 
\infaas tracks the dynamic status of variants (e.g., overloaded or interfered) using a state machine, to efficiently select the right variant for each query to meet the application requirements.  
Finally, \infaas combines VM-level (horizontal scaling) and \emph{model-level autoscaling} to dynamically react to the changing application requirements and request patterns. 
Given the large and complex search space of \modvars, we formulate an integer linear program (ILP) for our model-level autoscaling that finds the most cost-effective combination of \modvars, to meet the goals for queries in large scale inference serving.

Using query patterns derived from real-world applications and traces, we evaluate \infaas against existing inference serving systems, including Clipper~\cite{Clipper} and SageMaker~\cite{SageMaker}, with 175 variants generated from 22 model architectures, on AWS. 
Compared to Clipper, \infaas' ability to select suitable \modvars, leverage heterogeneous hardware (CPU, GPU, Inferentia), and share hardware resources across models and applications enables it to save 1.23$\times$ in cost, violate latency objectives 1.6$\times$ less often, and improve resource utilization by 2.8$\times$. 
At low load, \infaas saves cost by 21.6$\times$ compared to Clipper, and 21.3$\times$ compared to SageMaker.

\section{Challenges} \label{sec:motivation} 

\subsection{Selecting the right \modvar}
\label{sec:sel_modvar}

A \emph{\modvar} is a version of a model defined by the following aspects: 
(a) model architecture (e.g., ResNet50, VGG16), (b) programming framework, (e.g., TensorFlow, PyTorch, Caffe2, MXNet), (c) model graph optimizers (e.g., TensorRT, Neuron, TVM, XLA~\cite{tf-xla}), (d) hyperparameters (e.g., optimizing for batch size of 1, 4, 8, or 16), and (e) hardware platforms (e.g., Haswell or Skylake CPUs, V100 or T4 GPUs, FPGA, and accelerators, such as Inferentia~\cite{Inferentia}, TPU~\cite{TPU_ISCA}, Catapult~\cite{Catapult_ISCA}, NPU~\cite{alibaba-npu}).
Based on the 21 image classification models and the available hardware on AWS EC2~\cite{AWSEC2}, we estimate that the total number of possible \modvars would be \textbf{4,032}. 
The performance, cost, and accuracy trade-off space offered by these variants is large~\cite{mlperf,bianco2018benchmark}. 
As new inference accelerators are introduced and new optimization techniques emerge, the number of \modvars will only grow.

Existing inference serving systems require developers to identify the \modvar that can meet diverse performance, accuracy, and cost requirements of applications. 
However, generating and leveraging these variants requires a substantial understanding of the frameworks, model graph optimizers, and characteristics of hardware architectures, thus limiting the variants an application developer can leverage. 
As shown in Table~\ref{tab:facemodel_example}, one can use the same face recognition model for several applications, but selecting the appropriate \modvar depends on the requirements of an application~\cite{mlperf}.

We argue that inference serving systems should automatically and efficiently select a \modvar for each query on behalf of developers to align with application requirements. %

\subsection{Reducing cost as load varies}
\label{sec:motivation-varying-load}
\vspace{-1mm}
\begin{table}[t!]
  \centering
  \small
  \resizebox{\linewidth}{!}{
  \begin{tabular}{lccc}
  \toprule
  Variant (hardware, framework)  & Lat. (ms) & Req/s   & Cost (\$/s) \\ \midrule
  \rowcolor[HTML]{CBCEFB} 
  A (4 CPUs, TensorFlow)             & 200          & 5       & 1           \\
  B (1 Inferentia core, Neuron)      & 20           & 100     & 3         \\
  \rowcolor[HTML]{CBCEFB} 
  C (1 V100 GPU, TensorRT)             & 15           & 800     & 16           \\ \bottomrule
  \end{tabular}
  } %
  \caption{
    Latency, saturation throughput, and normalized cost (based on AWS pricing) for three ResNet50 variants.
  }
  \label{tab:example-variants}
\end{table}

\begin{table}[t!]
  \centering
  \small
  \resizebox{\linewidth}{!}{
  \begin{tabular}{ll|ccc|c}
\toprule
QPS  & SLO (ms) & \#Var. A & \#Var. B & \#Var. C & Cost (\$/s) \\ \midrule
\rowcolor[HTML]{CBCEFB} 
10   & 300      & 2         & 0         & 0         & 2    \\
10   & 50       & 0         & 1         & 0         & 3    \\
\rowcolor[HTML]{CBCEFB} 
1000 & 300      & 0         & 2         & 1         & 22   \\ \bottomrule
\end{tabular}
} %
  \caption{%
    Cheapest configuration (in \#instances) of variants from Table~\ref{tab:example-variants} to meet the QPS and SLO; last column shows total cost.
  }
  \label{tab:example-workload}
\end{table}

Query patterns and service level objectives (SLOs) of applications, such as real-time translation and video analytics, can vary unpredictably~\cite{AppliedML_FB, NoScope, videostorm}.
Provisioning for peak demand leads to high cost, hence distributed inference serving systems need to dynamically respond to changes. 
Traditional autoscaling focuses on horizontal virtual machine replication (VM-scaling), adding or removing worker machines~\cite{Autoscale,gujarati2017swayam}. %
Since inference serving is embarrassingly parallel, increasing the number of workers results in proportional increases in throughput and cost. 
However, relying only on worker replication may incur significant latency, as new machines must be spawned. 

Autoscalers used by existing inference serving systems~\cite{SageMaker, gujarati2017swayam,CloudML} replicate a statically-fixed (developer-specified) \modvar for all the queries of an application.
This is insufficient because: (a) the right variant may change with load (e.g., a CPU variant may be more suitable at low QPS to meet the cost SLOs) and (b) the hardware resources needed to replicate the same variant may not always be available (e.g., shortage of GPU instances at some point). 

In addition to using VM-scaling and replication-based model-scaling, we introduce \emph{model-level vertical scaling}, where we switch to a differently optimized variant as load changes.
The challenge is to identify which variant to scale to given available hardware resources and query requirements. 
Consider the example shown in Table~\ref{tab:example-variants} with three ResNet50 variants. 
Each variant runs on a different hardware resource and differs in latency, saturation throughput, and cost.
In Table~\ref{tab:example-workload}, we present three input loads and SLO requirements, with the goal of scaling to the most cost-effective combination of variants.
In the first case (QPS = 10 and SLO = 300ms), though all variants can meet the QPS and SLO, using two instances of Variant A is the cheapest choice (8$\times$ cheaper than using Variant C). 
In the second case, the load remains unchanged, but due to the stricter SLO, Variant B becomes the cheapest choice (5.3$\times$ cheaper than using Variant C). 
In the third case, the combination of two instances of Variant B and one of Variant C is the cheapest. 
This configuration is 1.45$\times$ cheaper than using two instances of Variant C (next cheapest configuration), and 9$\times$ cheaper than using 200 instances of Variant A (the most expensive configuration). 
Deciding the best configuration becomes more challenging as the number of variants increases and resource availability changes.

\subsection{Improving utilization at low load}
\label{sec:multi_tenancy}
\vspace{-1mm}
For predictable performance, one may serve each \modvar on a dedicated machine to exclusively access hardware resources.
But, this often results in underutilized resources and cost-inefficiency, especially at low load. 
Instead, the serving systems should support multi-tenancy by \emph{sharing resources across applications and models}, thereby improving utilization and the overall cost. 
Recent work~\cite{Gandiva, Tiresias} has shown the benefits of sharing GPUs for deep-learning training jobs.
ML inference jobs are typically less demanding for compute and memory resources than the training jobs, making inference jobs ideal for sharing GPUs and other accelerators~\cite{Salus, space_time_gpu}.

\begin{figure}[t]
\centering
    \begin{subfigure}[t]{0.55\linewidth}
        \centering
        \includegraphics[width=1.0\linewidth]{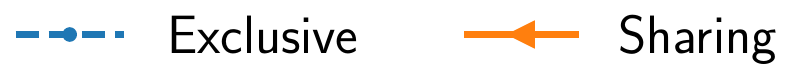}
    \end{subfigure}
    \\ \vspace{-1.2mm}
    \begin{subfigure}[t]{0.49\linewidth}
        \centering
        \includegraphics[width=1.0\linewidth]{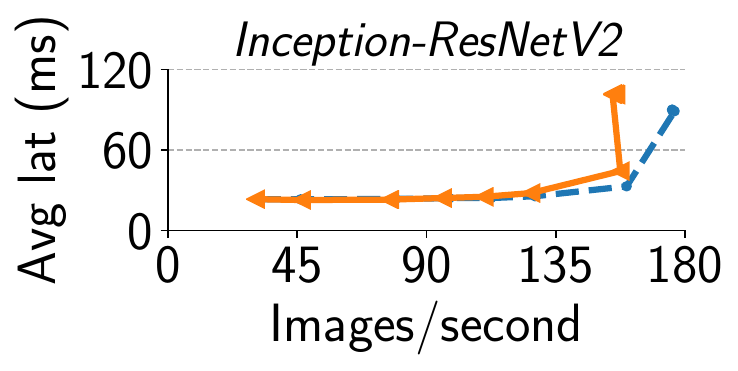}
        \label{fig:gpu_sharing_inception}
    \end{subfigure}
    \hfill
    \begin{subfigure}[t]{0.49\linewidth}
        \centering
        \includegraphics[width=1.0\linewidth]{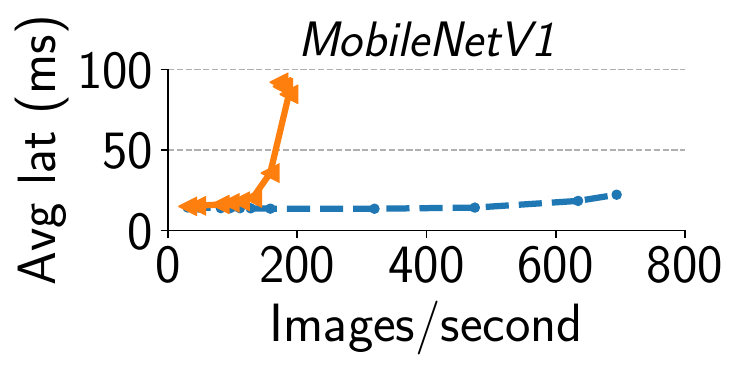}
        \label{fig:gpu_sharing_mobilenet}
    \end{subfigure}
    \vspace{-4mm}
    \caption{
     Impact of co-locating Inception-ResNetV2 and MobileNetV1 on a V100
     GPU. Both variants are TensorRT, batch-1, FP16.
     Graphs show average latency and throughput for each model running alone vs sharing, subjected to the same (QPS). 
    }
    \label{fig:gpu_sharing}
\end{figure}

However, how to share accelerators across multiple tenants while maintaining predictable performance is not obvious.
Figure~\ref{fig:gpu_sharing} shows the result of co-locating a large (Inception-ResNetV2) and a small (MobileNetV1) model on a GPU. 
At low load, sharing a GPU does not affect the performance of either model. 
At higher load, this co-location heavily impacts the performance of the small model, while the large model remains unaffected.
The point when co-location starts affecting the performance varies across models, and depends on both the load and the hardware architecture. 

An additional opportunity to improve resource utilization is to multiplex resources for online (realtime) and offline (batched) inference queries. 
Typical offline jobs, such as historical data analysis~\cite{polyzotis2017data} and image labeling~\cite{jing2015visual}, process a large number of queries and are latency tolerant (about minutes to hours). 
Thus, during periods of low or medium online load, offline and online jobs can run on the same machine. 
The trade-off lies in maximizing the resources used by offline jobs while minimizing the interference to online jobs~\cite{heracles}.

\section{\binfaas} \label{sec:overview-api} 
\paragraph{Design principles.} 
We design \infaas based on the following guidelines. 
First, \infaas should support a declarative API: Developers should not need to specify the model, model optimizations, suitable hardware platforms, or autoscaling configurations; they should only focus on high-level performance, cost, or accuracy requirements.
Across queries, the most suitable variant may differ; moreover, the best variant for the same query can be different based on the load and the status of variants and resources.
Thus, the second desirable property is that the system should automatically and efficiently select a \modvar, while considering the dynamic state of the \modvars and the hardware resources, for (a) serving each query, and (b) deciding how to scale in reaction to changing application load. 
Third, to improve resource utilization, the system should share hardware resources across \modvars and applications, without violating performance\-/cost constraints.
Finally, the system design should be modular and extensible to allow new \modvar selection policies. %
By following these design principles, we naturally address the challenges raised in Section~\ref{sec:motivation}.

\vspace{-2mm}
\paragraph{Functionality.}
\infaas generates new \modvars from the models registered by developers, and stores them in a repository (Section~\ref{subsec:arch}).
These variants are optimized along different dimensions using model graph optimizers such as Neuron and TensorRT.
For each inference query, \infaas automatically selects a \modvar to satisfy its performance, cost, and accuracy objectives (Section~\ref{subsec:select-modvar}).
\infaas' autoscaler combines VM-level autoscaling with model-level horizontal and vertical autoscaling to meet application performance and cost requirements while improving utilization of resources (Section~\ref{sec:autoscaling}).
\infaas introduces \emph{model-vertical autoscaling} that, through model selection, upgrades or downgrades to a differently optimized model-variant by leveraging the diversity of \modvars (Section~\ref{sec:modautoscale}).
\infaas efficiently maintains static and dynamic profiles of \modvars and hardware resources to support low latencies for selecting and scaling \modvars (Sections~\ref{subsec:arch} and~\ref{sec:modvar-selection}).
Finally, \infaas features the \modvar selection policy described in Section~\ref{sec:modvar-selection}, but allows developers to extend and customize it. %

\begin{table}[t]
  \centering
  \small
  \resizebox{0.95\linewidth}{!}{%
  \begin{tabular}{@{}p{2cm} p{6cm}@{}}\hline
  API                        & Parameters         \\ \hline
  \texttt{register\_model}    & modelName, modelBinary, valSet, appID    \\ \hline
  \texttt{online\_query}      & input(s), appID, latency, accuracy               \\ \hline
  \texttt{online\_query}      & input(s), modelName                                 \\ \hline
  \texttt{offline\_query}     & inputPath, outputPath, appID, accuracy    \\ \hline
  \texttt{offline\_query}     & inputPath, outputPath, modelName    \\ \hline
  \end{tabular}
  } %
  \caption{\infaas' declarative developer API. }%
  \label{tab:api_table}
\end{table}

\subsection{Model-less interface for inference} \label{subsec:interface} 
\vspace{-1mm} 
Table~\ref{tab:api_table} lists \infaas' model-less API.

\vspace{-1mm}
\paragraph{Model registration.}
Developers register one or more models using the \texttt{register\_model} API. 
This API accepts a developer-assigned model identifier (\texttt{modelName}), the model (\texttt{modelBinary}) in serialized format (e.g., a TensorFlow SavedModel or model in ONNX format), and a developer-assigned application identifier (\texttt{appID}).
Models for different prediction tasks within the same application (e.g., optical character recognition and language translation) can be registered with separate \texttt{appID}s. 
Lines 1-2 in Figure~\ref{fig:infaas-code} show how a developer registers two models, a ResNet50 and a MobileNet, for an application with \texttt{appID=detectFaceApp}. 
\infaas generates multiple variants from these already trained models. For instance, using ResNet50 alone, \infaas can generate about 50 variants by changing the batch size, the hardware, and the model graph optimizer (Section~\ref{subsec:arch}).
Note that \infaas is an inference serving system and does not train new models; \infaas only generates variants from already-trained models.
The \texttt{register\_model} API takes a validation dataset (e.g., \texttt{valSet}) as input to calculate the accuracy of the newly generated variants.
For each incoming query, \infaas automatically selects the right \modvar to meet the specified goals.

\vspace{-1mm}
\paragraph{Query submission.} %

Being declarative, \infaas' API allows developers to specify high-level goals without needing to specify the variants for their queries.
\infaas' API allows developers to submit online and offline queries in two ways: %
\begin{itemize}[leftmargin=*]
\item \emph{Specifying application requirements. }
Developers may submit queries for their application and specify high-level application performance, cost, and accuracy requirements (e.g., Line 3 in Figure~\ref{fig:infaas-code}). 
\infaas then navigates the search space of \modvars for the given application, and selects \modvars and scaling strategies. %
For instance, for a query with \texttt{appID=detectFaceApp}, \infaas searches for a suitable variant of ResNet50 and MobileNet to meet the goal of latency (200ms) and accuracy (above 70\%).

\item \emph{Specifying a registered model. }
Developers may use this interface to specify the model, \texttt{modelName}, they registered for the corresponding application (e.g., "ResNet50" for \texttt{detectFaceApp}). 
This interface supports developers who want direct control over the \modvar used. %
This is the only option offered by existing inference systems. 
\infaas then dynamically manages resources for the specified \modvar based on the observed workload.

\end{itemize} 

\begin{figure}[t] 
  \begin{lstlisting}[linewidth=\columnwidth,language=Python,basicstyle=\scriptsize\sffamily\mdseries,numberstyle=\tiny,numbersep=-1em,tabsize=2,aboveskip=0pt,belowskip=0pt]
  register_model("ResNet50",ResNet50.pt,valSet,detectFaceApp)
  register_model("MobileNet",MobileNet.pt,valSet,detectFaceApp)\end{lstlisting}
  \begin{lstlisting}[numbers=none,linewidth=\columnwidth,language=Python,basicstyle=\scriptsize\sffamily\mdseries,numberstyle=\tiny,numbersep=-1em,tabsize=2,aboveskip=0pt,belowskip=0pt]
 # Developer registered 2 models for the detectFaceApp;
 # INFaaS generates variants from these two registered models\end{lstlisting}
  \begin{lstlisting}[firstnumber=3,linewidth=\columnwidth,language=Python,basicstyle=\scriptsize\sffamily\mdseries,numberstyle=\tiny,numbersep=-1em,tabsize=2,aboveskip=0pt,belowskip=0pt,keywordstyle=\scriptsize\sffamily\mdseries]
  online_query(input.jpg,detectFaceApp, 200ms, 70%)\end{lstlisting}
  \begin{lstlisting}[numbers=none,linewidth=\columnwidth,language=Python,basicstyle=\scriptsize\sffamily\mdseries,numberstyle=\tiny,numbersep=-1em,tabsize=2,aboveskip=0pt,belowskip=0pt]
 # Developer submitted a query with "input.jpg" as the input
 # and specified requirements; INFaaS then selects a variant
 # automatically to meet 200ms latency and accuracy > 70%
\end{lstlisting} 
  \caption{Registering models and submitting queries with \infaas. } %
  \label{fig:infaas-code}
\end{figure}

\begin{figure}[tb!]
 \centering
 \includegraphics[width=\linewidth]{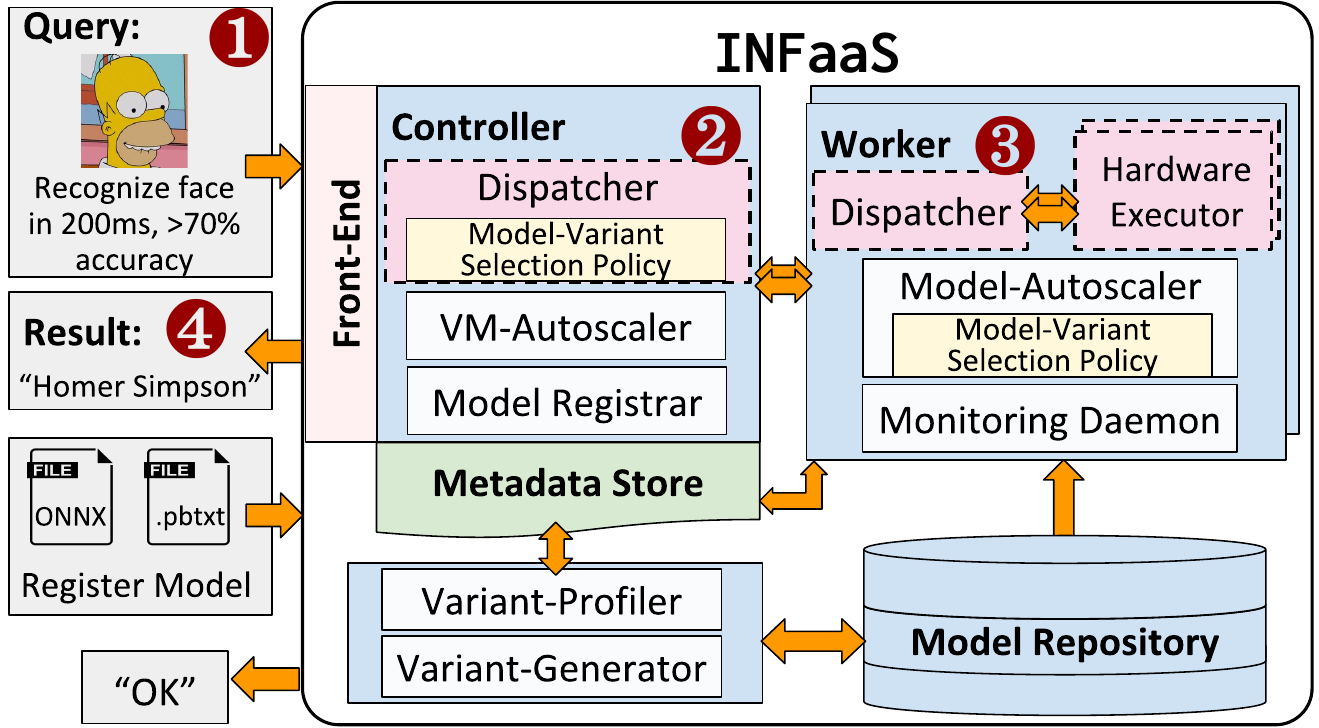}
 \caption{Architecture of \infaas.
  Modules in boxes with dashed border are on the critical path of serving queries.
  All other modules do not impact serving latency. 
  Numbered circles correspond to the typical life-cycle of queries. 
 }
 \label{fig:infaas_arch}
\end{figure}

\vspace{-1mm} 
\paragraph{\binfaas' serving workflow for an inference query. } 
Applications interact with \infaas by submitting inference queries through the \emph{Front-End}, logically hosted at the \emph{Controller} (see steps marked in Figure~\ref{fig:infaas_arch}). 
The Controller selects a \modvar and dispatches the inference query to a \emph{Worker} machine.
Worker machines further send inference queries to the appropriate \emph{Hardware Executors} according to the selected \modvar, and reply with inference results to applications.
Offline inference queries are served asynchronously; \infaas notifies the application when the offline job completes.

\subsection{Architecture} \label{subsec:arch} 
\vspace{-1mm} 
We now describe \infaas' architecture (Figure~\ref{fig:infaas_arch}) in detail. 

\noindent\textbf{Controller. } The logically-centralized controller receives model registration and inference requests. 
The controller hosts three modules: (a) The Dispatcher that uses the \modvar selection policy for selecting a variant to serve a query, 
(b) The VM-Autoscaler that is responsible for scaling the number of workers up and down based on the current load and resource utilization, and
(c) The Model Registrar that handles model registration.
For fault-tolerance, the controller is replicated using existing techniques~\cite{gujarati2017swayam,ConfluxDB}.

\noindent\textbf{Workers. }
Worker machines execute inference queries assigned by the controller. %
Hardware-specific Executor daemons manage the deployment and execution of \modvars.
The Dispatcher forwards each query to a specific \modvar instance through the corresponding hardware executor.
The Model-Autoscaler detects changes in the load and decides a scaling strategy that either replicates running variants or selects a different variant, within the worker. 
It uses the \modvar selection policy to select a variant to scale to. 
The Monitoring Daemon tracks the utilization of machines and variants, and manages resources shared by multiple variants to avoid SLO violations; it also decides when to process or pause offline requests to avoid interference with online serving. 

\noindent\textbf{Variant-Generator and Variant-Profiler. } 
From the registered models, the Variant-Generator generates %
\modvars optimized for different batch sizes, hardware, and hardware-specific parameters (e.g., number of cores on Inferentia) using model graph optimizers, including TensorRT~\cite{tensorRT} and Neuron~\cite{aws_neuron}. 
\infaas uses the validation set submitted by the developer to calculate the accuracy of the newly generated variants; \infaas records this information in the Metadata Store. 
The Variant-Generator does not train or produce new model architectures: variants are generated only from registered models. 
To help \modvar selection, the Variant-Profiler conducts one-time profiling for each variant where it measures statistics, such as the loading and inference latencies, and peak memory utilization. 
These parameters, along with the corresponding \texttt{appID}, accuracy, and maximum supported batch size are recorded in the Metadata Store.

\noindent\textbf{Model-Variant Selection Policy. } 
\infaas invokes the \modvar selection policy on two events. \newline
\emph{(Case I) On arrival of a query:} The controller's Dispatcher uses the policy to select a variant for each incoming query. 
This \modvar selection lies on the critical path of serving each query.
To reduce the latency of decision-making, we designed an efficient variant search algorithm (Section~\ref{subsec:select-modvar}). 

\noindent\emph{(Case II) On changes in query load:} %
As the query load changes, the worker's Model-Autoscaler uses the policy to determine whether to replicate existing variants, or vertically scale to a different variant.
The Model-Autoscaler monitors the incoming query load and the current throughput of \infaas to detect the need for scaling. 
If a change is detected, the Model-Autoscaler invokes the policy in the background to select a suitable scaling strategy (Section~\ref{sec:autoscaling}).

To allow for other \modvar selection algorithms, we designed \infaas to decouple policies from mechanisms~\cite{hydra_policy}. 

\noindent\textbf{Metadata Store. }
The Metadata Store enables efficient access to the static and dynamic data about workers and \modvars; this is needed for making \modvar selection and scaling decisions. 
This data consists of (a) the information about available model architectures and their variants (e.g., accuracy and profiled inference latency), and (b) the resource usage and load statistics of variants and worker machines.
The Metadata Store strategically uses data structures to ensure low access latencies ($\sim O(1)$) for efficient decision-making.
The Metadata Store runs on the same machine as the controller to reduce access latencies for selecting variants.
Implementation and data structure details are described in Section~\ref{sec:implementation}.

\noindent\textbf{Model Repository. }
The Model Repository is a high-capacity persistent storage medium that stores serialized variants that are accessible to workers when needed to serve queries.

\section{Selecting and Scaling Model-Variants}
\label{sec:modvar-selection}

\infaas uses the \modvar selection policy in two cases:  
(I) On arrival of a query: \infaas' controller needs to select a variant for each query to meet an application's high-level requirements (Section~\ref{subsec:select-modvar}). 
This invocation of the selection policy lies on the critical path of inference serving. 
(II) On changes in query load: As the query load changes, \infaas' workers must decide whether to switch to a differently optimized variant (Section~\ref{sec:autoscaling}).
The worker invokes the selection policy off the critical path. 
\infaas provides an internal API, \texttt{getVariant}, for invoking model-variant selection policy. %

In both cases, \infaas needs to consider both the static and dynamic states of variants and available resources.
Only considering statically-profiled metadata is insufficient, since the following aspects can significantly impact the observed performance and cost: a selected variant (a) may not be loaded, hence we need to consider its loading latency, (b) may be already loaded but serving at its peak throughput, (c) may be already loaded but experiencing resource contention from co-located inference jobs, and (d) may not be loaded due to lack of resources required for that specific variant.
We next describe how \infaas tracks the dynamic state of \modvars, and then describe the policy used in the two cases. 

\begin{figure}[tb!]
 \centering
 \includegraphics[width=\linewidth]{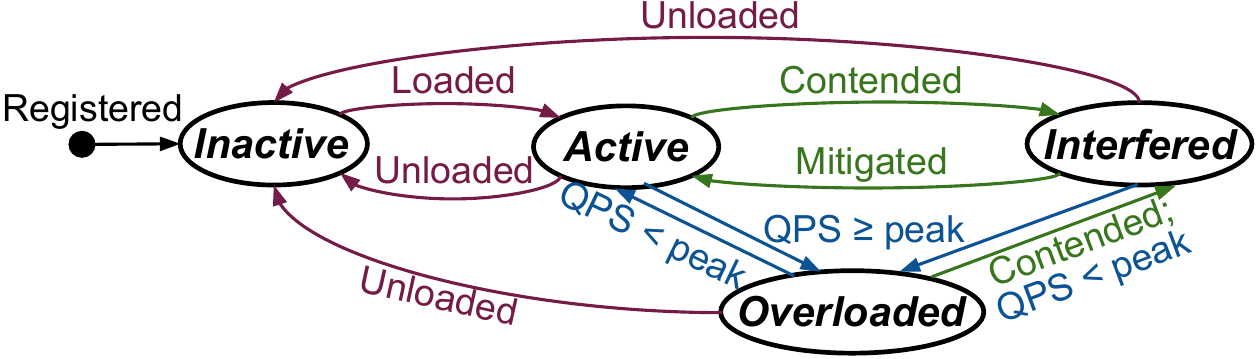}
 \caption{State machine capturing the dynamically changing status of \modvars. } %
 \label{fig:modvar-state}
\end{figure}

\noindent\textbf{State machine for the lifecycle of \modvars. }
To track the dynamic state of each \modvar instance per-application, \infaas uses a state machine (shown in Figure~\ref{fig:modvar-state}). 
All the registered and generated \modvars start in the \emph{Inactive} state: they are not loaded on any worker. 
Once a variant instance is loaded, it transitions to the \emph{Active} state.
These variant instances are serving less than their peak throughput, tracked by the worker's monitoring daemons.
Variant instances enter the \emph{Overloaded} state when they serve at their peak throughput.
Finally, variant instances in the \emph{Interfered} state are not overloaded but are still experiencing higher inference latencies than the profiled values.
Interference occurs when co-located variants contend over shared resources (e.g., caches, memory bandwidth, or hardware threads).

\vspace{-1mm}
\paragraph{Maintaining the state machine.}
The state machine for each \modvar instance is maintained by the worker's monitoring daemons and is organized in the Metadata Store for fast access.
This enables \infaas' Dispatcher to use the \modvar selection policy for serving queries on the order of hundreds of $\mu$s to ms (assessed further in Section~\ref{sec:exp-decisionmaking}). 
State machine implementation details are described in Section~\ref{sec:implementation}.

\subsection{Case I: On arrival of a query} \label{subsec:select-modvar} 
\vspace{-1mm}
When a query arrives, \infaas' Dispatcher invokes the \texttt{getVariant} method of \modvar selection policy to choose a variant.
For this case, the input to \texttt{getVariant} is the query's requirements, and the output is the variant and worker to serve the query. 
\texttt{getVariant} first checks whether any variants in the \emph{Active} state match the query's requirements. 
Variants in \emph{Active} state do not incur a loading latency. 
If such a variant is found, \infaas dispatches the query to the least-loaded worker running the variant instance.
Otherwise, \infaas considers variants in the \emph{Inactive} state: \texttt{getVariant} first enquires the Metadata Store and retrieves the variant with the lowest combined loading and inference latency that matches the query's requirements. 
If such a variant is found, \infaas sends the query to the worker with the lowest utilization on the variant's target hardware.
Otherwise, since no registered or generated variant can meet the developer's requirements, \infaas suggests a variant that can achieve the closest target accuracy and/or latency. 
We assess the efficiency of this policy over brute-force search in Section~\ref{sec:exp-decisionmaking}.  

\vspace{-1mm}
\paragraph{Mitigating performance degradation. }
For better resource utilization, \infaas co-locates variants on hardware resources; as a result, they may interfere and cause SLO violations.
To prevent such violations, \infaas avoids selecting variants that are in the \emph{Interfered} or \emph{Overloaded} state. 
For interfered variants, \infaas triggers a mitigation process in the background to avoid affecting the performance of online serving. 
If there are idle resources available on the same worker, this mitigation process migrates the variant in the \emph{Interfered} state to the available resources (e.g., a different set of cores). 
This avoids the need to fetch a variant from the model repository. 
If no resources are available for loading the variant on the worker, the worker asks the controller's Dispatcher to place the variant on the least-loaded worker. 
For variants in the \emph{Overloaded} state, \infaas' Model-Autoscaler assesses whether it is more cost-effective to scale to a different variant (see Section~\ref{sec:modautoscale}).

\vspace{-1mm}
\paragraph{Extensibility.}
The state machine and \modvar selection policy are extensible.
For instance, \texttt{getVariant} can be extended to prioritize particular variants in the \emph{Active} state (e.g., prefer least  power-hungry variants).

\subsection{Case II: On changes in query load} \label{sec:autoscaling} 
\vspace{-1mm}

As query load changes, \infaas needs to revisit its variant selection decision to check whether a different variant is more cost-efficient. 
Existing inference serving systems~\cite{gujarati2017swayam,CloudML,Clipper,TensorFlowServing,SageMaker} are agnostic to the diversity of \modvars, and only replicate a statically fixed (developer-specified) \modvar for all the queries of an application.
However, as discussed in Section~\ref{sec:motivation-varying-load}, autoscaling that replicates the same \modvar alone is not enough because: (a) the right \modvar changes with load and (b) the required resources might not be available to replicate a specific variant. 

For \infaas' autoscaling, in addition to using traditional VM-level, horizontal autoscaling, we introduce \emph{model-vertical scaling}: change (upgrade or downgrade) to a differently optimized \modvar, thus leveraging the diversity of \modvars. 
\infaas' autoscaling is a joint effort between the workers and the controller. 
Each worker hosts a Model-Autoscaler that consults with the \modvar selection policy to make model-level autoscaling decisions (Sections~\ref{sec:modautoscale}, and~\ref{subsec:greedyheuristic}). 
The controller hosts a VM-Autoscaler that makes VM-level autoscaling decisions (Section~\ref{sec:master-autoscaler}).

\subsubsection{Model-Autoscaler at each worker} \label{sec:modautoscale} 
\vspace{-1mm}
To react to the changes in query load, \infaas' Model-Autoscaler needs to decide the type and number of \modvars to use while minimizing the cost of running the variants. %
To figure out the type and number of \modvars needed, we formulated the following integer linear program (ILP) that decides a scaling action (replicate, upgrade, or downgrade) for each variant. 
This ILP minimizes the total cost of scaling actions for all the variants to meet the incoming query load.

\noindent{\textbf{Formulation. }}
For an application, the outcome (optimization variable) of our ILP is the optimal scaling action, $\delta_{ij}$, for each \modvar $v_{ij}$, variant $j$ of model architecture $i$. 
$\delta_{ij}$ is an integer that captures the scaling action as follows: 
(a) A positive value denotes loading instances of the variant, (b) a negative value denotes unloading instances of this variant, and (c) a value of zero denotes no scaling needed. 
For a variant $v_{ij}$ that is already loaded, a positive value of $\delta_{ij}$ indicates a \emph{replicate} action. 
A positive value of $\delta_{ij}$ for a variant $v_{ij}$ that is not already loaded indicates an \emph{upgrade} or \emph{downgrade} action depending on the hardware cost of $v_{ij}$.

The objective function, that our ILP minimizes, is the total cost of all the chosen scaling actions. 
For a variant $v_{ij}$, this cost for an action $\delta_{ij}$ is the sum of the hardware cost (in \$/second), and the loading latency (in seconds) of the variant: 

\vspace{-3mm}
$$\mathrm{Cost}(\delta_{ij}) = C_{ij} ( \delta_{ij} + \lambda T^{\mathrm{load}}_{ij} \max(\delta_{ij}, 0))$$
where $C_{ij}$ is the hardware cost (in \$/second) for running the variant, $T^{\mathrm{load}}_{ij}$ is the loading latency of the variant, and $\lambda$ (in $\frac{1}{\text{second}}$) is a tunable parameter for the query load unpredictability. 
Large values of $\lambda$ place more weight on minimizing loading latency to meet SLOs when the query load is unpredictable or spiky.
Small values of $\lambda$ place more weight on minimizing the hardware cost when the query load is more stable. %

Thus, our objective function, the total cost for all the variants, is: $\sum_{i, j} \text{Cost}(\delta_{ij})$. 
We impose the following constraints on our ILP: \newline %
\noindent\textbf{(1)} With the chosen scaling actions, \infaas supports the incoming query load. \newline 
\noindent\textbf{(2)} The newly-loaded instances satisfy applications' SLOs. \newline 
\noindent\textbf{(3)} The resources consumed by all variants do not exceed the total system resources. \newline 
\noindent\textbf{(4)} The number of running instances is non-negative.

\noindent We write these constraints formally as:  
\vspace{-2mm}
\begin{equation*}
\begin{array}{llr}%
\sum_{i, j} Q_{ij}(N_{ij} + \delta_{ij})  \geq  L + \texttt{slack} & \text{for all } i, j &  (1) \\
T^{\mathrm{inf}}_{ij} \leq S & \text{if } \delta_{ij} > 0 & (2) \\
\sum_{i, j} R^{\mathrm{type}}_{ij}(N_{ij} + \delta_{ij}) \leq R^{\mathrm{type}}_{\mathrm{total}} & \text{for all types}  & (3) \\
N_{ij} + \delta_{ij} \geq 0  &  \text{for all } i, j  &  (4)
\end{array}
\end{equation*} 
where 
(a) $Q_{ij}$: the saturation QPS of variant $v_{ij}$, 
(b) $N_{ij}$: the number of running instances of variant $v_{ij}$, 
(c) $L$: the incoming query load, 
(d) \texttt{slack}: configurable headroom to absorb sudden load spikes, 
(e) $T^{\mathrm{inf}}_{ij}$: the inference latency of variant $v_{ij}$, 
(f) $T^{\mathrm{load}}_{ij}$: the loading latency of variant $v_{ij}$, %
(g) $S$: SLO of the considered application, %
(h) $R^{\mathrm{type}}_{ij}$: the resource requirements of variant $v_{ij}$, for a resource type (CPU cores, CPU memory, GPU memory, number of Inferentia cores), %
and (i) $R^{\mathrm{type}}_{\mathrm{total}}$: the total available amount of resources of a type (CPUs, GPUs, Inferentia cores) %
on the underlying worker machine. 

The \modvar selection policy queries the Metadata Store to get the values of these variables. 

\vspace{-1mm} 
\paragraph{Practical limitation of the ILP. } 
Unfortunately, this ILP is NP-complete and hence offers limited practical benefits~\cite{np-completeness,von1978bound,li2013energy}. 
Our ILP formulation has to exhaustively search through all the \modvars, 
track their dynamically changing state, and accurately estimate the QPS each variant can support to find a scaling configuration that can sustain the changed query workload.  
Gurobi~\cite{gurobi} took 23 seconds to find the optimal number of running variant instances across 50 model architectures, and 50 seconds for 100 model architectures. 
To meet realtime requirements of latency-sensitive applications, \infaas must have sub-second response time to query workload changes. 

\subsubsection{A Greedy Heuristic} \label{subsec:greedyheuristic} 
\vspace{-1mm}
The time taken to solve each instance of our ILP makes it impractical to use for \infaas.
Instead, we design a greedy heuristic algorithm that replaces our ILP's large search space by a subset of \modvars.
This pruned search space allows \infaas to meet the outlined constraints at sub-second latency. 
We empirically evaluate the effectiveness of this algorithm in Section~\ref{sec:workerscale}.
Each worker machine runs a Model-Autoscaler that, together with the \modvar selection policy, approximates this ILP as follows: (a) Identify whether the constraints are in danger of being violated, (b) Consider two strategies, replicate or upgrade/downgrade, to satisfy the constraints, (c) Compute the objective for each of these scaling actions, and pick the one that minimizes the objective cost function, and (d) Coordinate with the controller to invoke VM-level autoscaling if constraints cannot be satisfied with model-level autoscaling.

\noindent{\textbf{Scaling up algorithm: }} 
To decide if there is a need to scale (Constraint \#1), the Model-Autoscaler estimates the current headroom in capacities of running \modvars, given the profiled values of their saturation throughput, and the current load they are serving. 
We compute the current load served by a variant using the batch size and number of queries served per second. 
The load served by a worker is estimated by summing the load served by all running variants.  
The saturation throughput of all running variants is estimated in a similar manner using the profiled values of \modvars. 
The Model-Autoscaler then computes the current headroom of a worker as the ratio of the combined saturation throughput and the combined load currently served by the running variants on that worker. 
\infaas maintains a minimum headroom, \texttt{slack-threshold}, on each worker to absorb sudden load spikes. 
We discuss the value of this tunable parameter in Section~\ref{sec:implementation}. 
When the current headroom is below the required minimum \texttt{slack-threshold}, the Model-Autoscaler concludes that we need to scale, and proceeds to answer the second question: \emph{how} to scale (replicate or upgrade) to meet the incoming query load. %

To decide how to scale, the Model-Autoscaler uses the \modvar selection policy's \texttt{getVariant} method to select the cheapest option between replication and upgrading.
For this case, the input to \texttt{getVariant} is the incoming query load, and the output is the set of scaling actions. %
The policy first estimates the cost of model-horizontal scaling (replication) by estimating the number of instances of the running variant that would be added to meet the incoming query load (Constraints \#1, \#4).
Secondly, the policy estimates the cost of model-vertical scaling (upgrade), by querying the Metadata Store to select variants of the same model architecture that can meet the SLO (Constraint \#2), and support a higher throughput than the currently running variant.
The required number of instances for these variants to meet the incoming query load is then estimated. 
Finally, the \modvar selection policy computes the cost function of our ILP, by using the hardware cost (\$/s) and the variant loading latency to decide whether to replicate the running variant, or upgrade to a variant that supports higher throughput. 
The available resources on the worker limit the number of variant instances it can run (Constraint \#3).
Thus, if the strategy requires more resources than are available on the current worker (e.g., hardware accelerator), the worker coordinates with the controller to load the variant on a capable worker. 

\noindent{\textbf{Scaling down algorithm: }} 
To decide if and how to scale down (remove replicas or downgrade), the Model-Autoscaler on each worker uses the \modvar selection policy that follows a similar algorithm explained above for scaling up. 
At regular intervals, this policy checks if the incoming query load can be supported by removing an instance of the running variant, or downgrading to a cheaper variant (optimized for a lower batch size or running on different hardware). 
The Model-Autoscaler waits for $T_v$ time slots before executing the chosen strategy for a variant $v$, to avoid scaling down too quickly. 
$T_v$ is set equal to the loading latency of variant $v$.

\vspace{-1mm}
\subsubsection{VM-Autoscaler at controller} \label{sec:master-autoscaler} 
\vspace{-1mm}

In addition to model-level scaling, \infaas also scales the worker machines for deploying variants. 
Following the mechanisms used in existing systems~\cite{burns2016borg,mesos,Clipper,SageMaker,CloudML}, the VM-Autoscaler decides when to bring a worker up/down: %

\begin{enumerate} 
\item When the utilization of any hardware resource exceeds a configurable threshold across all workers, the VM-Autoscaler adds a new worker with the corresponding hardware resource.
We empirically set the threshold to 80\%, considering the time to instantiate VMs (20-30 seconds): a lower threshold triggers scaling too quickly and unnecessarily adds workers; a higher value may not scale in time. %

\item When variants on a particular hardware platform (e.g., GPU) are in the \emph{Interfered} state across all workers, the VM-Autoscaler adds a worker with that hardware resource. 

\item When more than 80\% of workers have \emph{Overloaded} variants, the VM-Autoscaler starts a new worker. %
\end{enumerate} 
To improve utilization, \infaas dispatches requests to workers using an online bin packing algorithm~\cite{binpacking}.

\vspace{-1mm}
\section{Implementation} \label{sec:implementation} 
\vspace{-1mm}

We implemented \infaas in about 20K lines of C++ code\footnote{We will make source code available with the publication of this paper.} %
\infaas' API and communication logic between controller and workers are implemented using gRPC in C++~\cite{grpc}.
Developers can interact with \infaas by issuing gRPC requests in languages supported by gRPC, such as Python, Go, and Java. %
\infaas uses AWS S3~\cite{AWSS3} for its Model Repository. %
The \modvar selection policy is implemented as an extensible C++ library that is linked into the controller's Dispatcher and worker's Model-Autoscaler.
\texttt{getVariant} is a virtual method, and can be overridden to add new algorithms.

On the controller machine, the Front-End, Dispatcher, and Model Registrar are threads of the same process for efficient query dispatch.
The VM-Autoscaler is a separate process, that polls system status periodically.
We swept the polling interval between 0.5-5 seconds at 0.5 second increments (similar to prior work~\cite{mage,Pocket}), and arrived at a 2 seconds polling interval.
Longer intervals did not scale up fast enough, especially during load spikes, and shorter intervals were too frequent given VM start-up latencies.

On worker machines, the Dispatcher and monitoring daemon run as separate processes.
Every two seconds, the monitoring daemon updates compute and memory utilization of the worker, loading, and average inference latencies, along with the current state (as noted in Figure~\ref{fig:modvar-state}) for each variant running on that worker, to the Metadata Store. %
We deployed custom Docker containers for PyTorch and Inferentia variants, and leveraged Triton Inference Server-19.03~\cite{TRT_Server} to support TensorRT, Caffe2, and TensorFlow variants on GPU.
We used the TensorFlow Serving container for TensorFlow variants on CPU~\cite{TensorFlowServing}. 
The Model-Autoscaler's main thread makes scaling decisions periodically. 
We swept the same range as the VM-Autoscaler, and arrived at a 1 second polling interval.
The interval is shorter than the VM-Autoscaler's polling interval as model loading latencies are shorter than VM start-up latencies. %
The main thread also manages a thread pool for asynchronously loading and unloading \modvars.
To tune \texttt{slack-threshold}, we explored values between 1.01 and 1.1~\cite{jockey}, and set it to 1.05.
In our setup, lower thresholds did not scale variants fast enough to meet load changes, while higher thresholds scaled variants too quickly.

We built the Variant-Generator using TensorRT~\cite{tensorRT} and Neuron~\cite{aws_neuron}; it is extensible to other similar frameworks~\cite{TVM,Oh:2018}.
For each variant, the Variant-Profiler records the latency for batch sizes from 1 to 64 (power of two increments).
For natural language processing models, we record the latencies of varying sentence lengths for each of these batch sizes.

We built the Metadata Store using Redis~\cite{redis-www} and the Redox C++ library~\cite{Redox}. 
The Metadata Store uses hash maps and sorted sets for fast metadata lookups that constitute the majority of its queries. 
Per-application, each \modvar instance's state is encoded as a \{variant, worker\} pair that can be efficiently queried by the controller and worker.

\infaas, by design, automates the tedious decision-making aspects of inference serving, thereby allowing an intuitive interface for developers. %
When needed, we note that \infaas' thresholds are configurable, and we used the following values: (a) Resource utilization for pausing offline job (40\%), (b)  resource utilization for VM-Autoscaler's scale up action (80\%), and (c) Model-Autoscaler's \texttt{slack-threshold} (1.05).

\section{Evaluation} \label{sec:eval} 
\vspace{-1mm}

We first compare \infaas with all of its optimizations and features to existing systems (Section~\ref{sec:piat}).
To further demonstrate the effectiveness of \infaas' design decisions and optimizations, we evaluate its individual aspects: \modvar selection, scaling (Section~\ref{sec:workerscale}), and SLO-aware resource sharing (Section~\ref{sec:shareresources}). 
Finally, we quantify the overheads of \infaas' decision-making (Section~\ref{sec:exp-decisionmaking}).
We begin by describing the experimental setup common across all experiments, the baselines, the \modvars, and the workloads.

\begin{table}[t!]
  \resizebox{\linewidth}{!}{%
  \centering
 \footnotesize
 \begin{tabular}{@{}l|ccc@{}}
\toprule
Features                   & \begin{tabular}[c]{@{}l@{}}Clipper,\\ TFS, TIS\end{tabular} & \begin{tabular}[c]{@{}c@{}}SageMaker, \\ AI Platform\end{tabular} & \infaas                                 \\ \hline
Model-less abstraction         & \textcolor{red!70!black}{No}   & \textcolor{red!70!black}{No}       & \textcolor{green!50!black}{Yes}   \\
Variant selection              & \textcolor{red!70!black}{Static} & \textcolor{red!70!black}{Static}     & \textcolor{green!50!black}{Dynamic}\\
VM-autoscaling & \textcolor{red!70!black}{No}   & \textcolor{green!50!black}{Yes}       & \textcolor{green!50!black}{Yes}   \\
Model-horizontal scaling & \textcolor{red!70!black}{No}   & \textcolor{green!50!black}{Yes}       & \textcolor{green!50!black}{Yes}   \\
Model-vertical scaling   & \textcolor{red!70!black}{No}   & \textcolor{red!70!black}{No}       & \textcolor{green!50!black}{Yes}   \\
SLO-aware resource sharing     & \textcolor{red!70!black}{No}   & \textcolor{red!70!black}{No}       & \textcolor{green!50!black}{Yes}   \\ \bottomrule
\end{tabular}
  } %
  \caption{Comparison between \infaas and the baselines.}
  \label{tab:compare-baselines}
\end{table}

\noindent\textbf{Experimental setup.} 
We deployed \infaas on a heterogeneous cluster of AWS EC2~\cite{AWSEC2} instances.
We hosted the controller on an \texttt{m5.2xlarge} instance (8 vCPUs, 32GiB DRAM), and workers on \texttt{inf1.2xlarge} (8 vCPUs, 16GiB DRAM, one AWS Inferentia), \texttt{p3.2xlarge} (8 vCPUs, 61GiB DRAM, one NVIDIA V100 GPU), and \texttt{m5.2xlarge} instances.
All instances feature Intel Xeon Platinum 8175M CPUs operating at 2.50GHz, Ubuntu 16.04 with 4.4.0 kernel, and up to 10Gbps networking speed.

\noindent{\bf{Baselines. }} 
To the best of our knowledge, no existing system provides a model-less interface like \infaas;
state-of-the-art serving systems require developers to specify the variant and hardware.
For a fair comparison, we configured \infaas to closely resemble the resource management, autoscaling techniques, and APIs of existing systems, including TensorFlow Serving~\cite{TensorFlowServing} (TFS), Triton Inference Server (TIS)~\cite{TRT_Server}, Clipper~\cite{Clipper}, AWS SageMaker~\cite{SageMaker} (SM), and Google AI Platform~\cite{CloudML}.
Specifically, we compared \infaas to the following baseline configurations for online query execution:

\begin{itemize}
\item \textbf{\static}:
Derived from TFS, TIS, and Clipper, this baseline pre-loads \modvars, and requires developers to set a pre-defined number of variant instances.
Thus, we set the number of variant instances such that \static achieves the highest performance given available resources.

\item \textbf{\indivscale}: Derived from SageMaker and AI Platform, this baseline scales each \modvar horizontally, but does not support model-vertical scaling that \infaas introduces.
\end{itemize} 

\noindent 
Table~\ref{tab:compare-baselines} lists the differences between baselines and \infaas.
Configuring the baselines with \infaas (a) allowed for a fair comparison by removing variabilities in execution environments (e.g., RPC and container technologies), and (b) enabled us to evaluate our design decision individually by giving the baselines access to \infaas' features and optimizations, including: support for model graph optimizations, and \infaas' detection and mitigation of variant performance degradation.

\vspace{-1mm}
\paragraph{\vanillaclipper vs \static. }
To validate our baseline configurations through \infaas, we evaluated \static against the open-source Clipper deployment (\vanillaclipper)~\cite{clipper_github} with its adaptive batching and prediction caching features enabled.
We deployed two ResNet50 TensorFlow CPU instances for each.
For \vanillaclipper, we swept its adaptive batching SLO from 500ms to 10 seconds, and found it achieved its maximum throughput (7 QPS) when setting the SLO to 1 second.
For the same SLO, \static was able to achieve 10 QPS.
As prior work has noted~\cite{nexus}, \vanillaclipper's adaptive batching is insufficient for maintaining a high QPS, because it relies on an external scheduler to allocate resources for it.
Since \static benefits from \infaas' resource allocation and management, variant performance degradation detection and mitigation, and variant optimizations, we use \static in the place of \vanillaclipper for the remainder of our evaluation.

\paragraph{\vanillasagemaker vs \indivscale. }
We also validated that the latency and throughput of CPU, GPU, and Inferentia variants with \indivscale closely match \vanillasagemaker, while offering the benefits outlined for \static.
Thus, we use \indivscale in the place of \vanillasagemaker as our baseline.

\begin{table}[t]
  \resizebox{\linewidth}{!}{%
  \centering
  \begin{tabular}{l|l|l|l}
  \hline
  \textbf{Model Family (Task)} & \textbf{\#Vars} & \textbf{Model Family (Task)} & \textbf{\#Vars} \\ \hline
  MobileNet (classification)  & 13          & VGG (classification)        & 30          \\ \hline
  AlexNet (classification)    & 9           & Inception (classification)  & 25          \\ \hline
  DenseNet (classification)   & 22          & NasNet (classification)     & 6           \\ \hline
  ResNet (classification)     & 61          & GNMT (translation)     & 9           \\ \hline
  \end{tabular}
  } %
  \caption{\small Model architectures, tasks, and associated variants.}
  \label{tab:mod_arch_vars}
\end{table}

\noindent{\bf{\Modvars. }}
Guided by the MLPerf Inference benchmark~\cite{mlperf}, we collected a variety of models.
Table~\ref{tab:mod_arch_vars} shows the 8 model families (22 architectures) and the number of associated variants.\footnote{See Appendix A for more details of the model architectures used. }
Our models are pre-trained using Caffe2, TensorFlow, and PyTorch.
Our classification models are pre-trained on ImageNet~\cite{Imagenet}; translation models are pre-trained on the WMT16~\cite{WMT16} English-German dataset.
We generated 175 variants in total, differing in the frameworks (Caffe2, TensorFlow, PyTorch), compilers (TensorRT, Neuron), batch sizes (1 to 64), and hardware platforms (CPU, GPU, Inferentia).

\noindent{\bf{Workloads. }} 
We evaluated using both synthetic and real-world application query patterns.
For synthetic workloads, we used common patterns~\cite{Quasar} indicating flat and fluctuating loads, with a Poisson inter-arrival rate~\cite{mlperf,gupta2020deeprecsys}.
For real-world inference workloads, we used the timing information from a Twitter trace from 2018 collected over a month~\cite{twitter} since there is no publicly available inference serving production traces.
Twitter queries are likely passed through hate speech detection models~\cite{davidson2017automated} before being posted.
Furthermore, as noted in recent work on inference serving~\cite{mark}, this trace resembles real inference workloads with both diurnal patterns and unexpected spikes (consistent with production serving workloads~\cite{shahrad2020serverless}).
For each experiment, we randomly selected one day out of the month from the Twitter trace.

\subsection{\binfaas with production workload} \label{sec:piat}
\vspace{-1mm}

We now show that through model selection, resource allocation, and autoscaling mechanisms, \infaas improves the throughput, cost, utilization, and reduces SLO violations compared to the baselines. %

\noindent\textbf{Experimental setup. } 
We mapped the Twitter trace to a range between 10 and 1K QPS for a total of 113,420 batch-1 queries.
We used all 22 model architectures. 
Based on prior work~\cite{pretzel}, we used a Zipfian distribution for model popularity. 
We designated 4 model architectures (DenseNet121, ResNet50, VGG16, and InceptionV3) to be \emph{popular} with 50ms SLOs and share 80\% of the load.
The rest are \emph{cold} models with SLO set to 1.5$\times$ the profiled latency of each model's fastest CPU variant.
Baselines statically selected GPU variants for popular models, and CPU variants for the rest; they used 5 CPU and 7 GPU workers.
\infaas started with 5 CPU, 5 GPU, and 2 Inferentia workers.
We computed the worker costs based on AWS EC2 pricing~\cite{EC2_price}. %

\begin{figure}[t]
\centering
    \begin{subfigure}[t]{0.99\linewidth}
        \centering
        \includegraphics[width=1.0\linewidth]{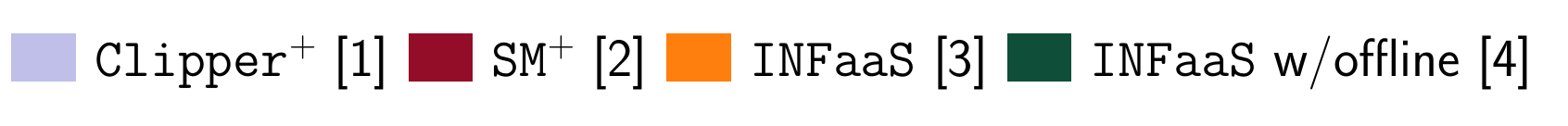}
    \end{subfigure}
    \vspace{-1mm}
    \\
    \begin{subfigure}[t]{0.49\linewidth}
        \centering
        \includegraphics[width=1.0\linewidth]{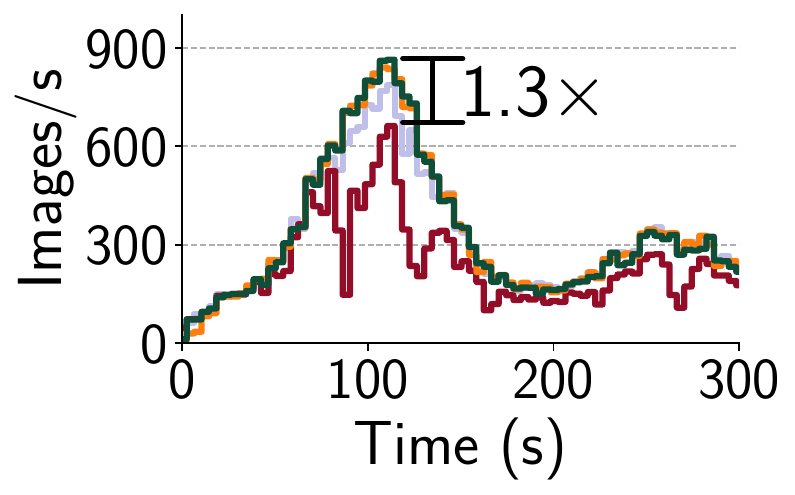}
        \label{fig:mixedload_online_qps}
    \end{subfigure}
    \hfill
    \begin{subfigure}[t]{0.49\linewidth}
        \centering
        \includegraphics[width=1.0\linewidth]{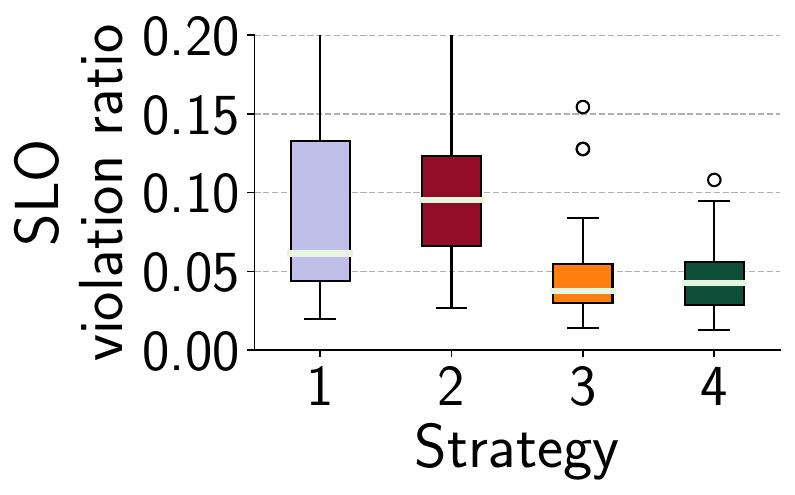}
        \label{fig:mixedload_online_slo}
    \end{subfigure}
    \vspace{-5mm}
    \caption{
     Throughput and SLO violation ratio (\# of SLO violations by total \# of queries), measured every 4 seconds.
       Each box shows the median, 25\% and 75\% quartiles; whiskers extend to the most extreme non-outlier values (1.5$\times$ interquartile range).
       Circles show the outliers. %
    }
    \label{fig:mixedload_online}
\end{figure}

\noindent\textbf{Results and discussion.} 
Figure~\ref{fig:mixedload_online} shows that \infaas achieved 1.1$\times$ and 1.3$\times$ higher throughput, and 1.63$\times$ and 2.54$\times$ fewer SLO violations compared to \static and \indivscale, respectively.
\infaas scaled models both horizontally and vertically: it upgraded to Inferentia or GPU (higher batch) variants when needed.
In reaction to the increased load, \infaas added a 3\textsuperscript{rd} Inferentia worker at 40 seconds.
Although \indivscale scales variants horizontally, it achieved lower throughput and violated more SLOs due to frequently incurring variant loading penalties and being unable to upgrade variants.  
By leveraging variants that span heterogeneous hardware (CPU, GPU, Inferentia), \infaas achieved 1.23$\times$ lower cost, while keeping SLO violations under 4\% on average.
\infaas also load-balanced requests and mitigated overloaded or interfered variants.
This resulted in an average worker utilization of 48.9\%, with an average GPU DRAM utilization of 58.6\% (5.6$\times$ and 2.8$\times$ higher than \indivscale and \static, respectively).

We then added 4 concurrent offline requests to evaluate the efficiency of \infaas' resource management.
Each offline request contained 500 input images and %
specified the ResNet50 PyTorch variant. 
As shown in Figure~\ref{fig:mixedload_online}, \texttt{\infaas w/offline} maintained similar throughput and SLO violations compared to \infaas only serving online requests.
Across 3 runs, 756 images on average were processed by offline queries.
By dynamically throttling offline requests, \infaas guaranteed SLOs for online requests.
\emph{\infaas achieved higher performance (1.3$\times$ higher throughput) and resource utilization (5.6$\times$ higher GPU utilization), and lower SLO violations (2$\times$ lower) and cost (1.23$\times$ lower) compared to the baselines.}

\vspace{-1mm}
\subsection{Selecting and scaling \modvars} \label{sec:workerscale}
\vspace{-1mm} 
\begin{figure}[t]
\centering
    \begin{subfigure}[t]{0.99\linewidth}
        \centering
        \includegraphics[width=\linewidth]{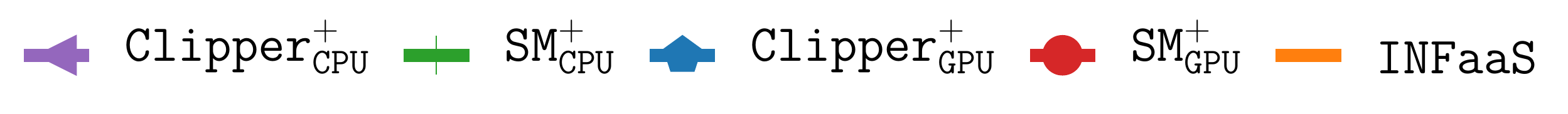}
    \end{subfigure}
    \vspace{-1mm}
    \\
    \begin{subfigure}[t]{.32\linewidth}
        \centering
        \includegraphics[width=1.0\linewidth]{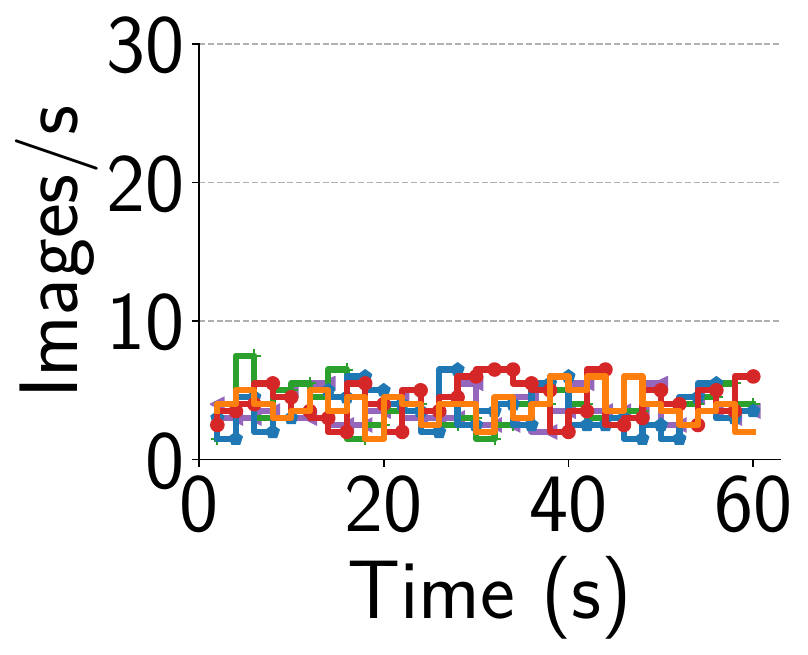}
        \vspace{-6mm}
        \caption{Flat, low load}
        \label{fig:exp-autoscaling-a}
    \end{subfigure}
    \hspace{-1.5mm}%
    \begin{subfigure}[t]{.345\linewidth}
        \centering
        \includegraphics[width=1.0\linewidth]{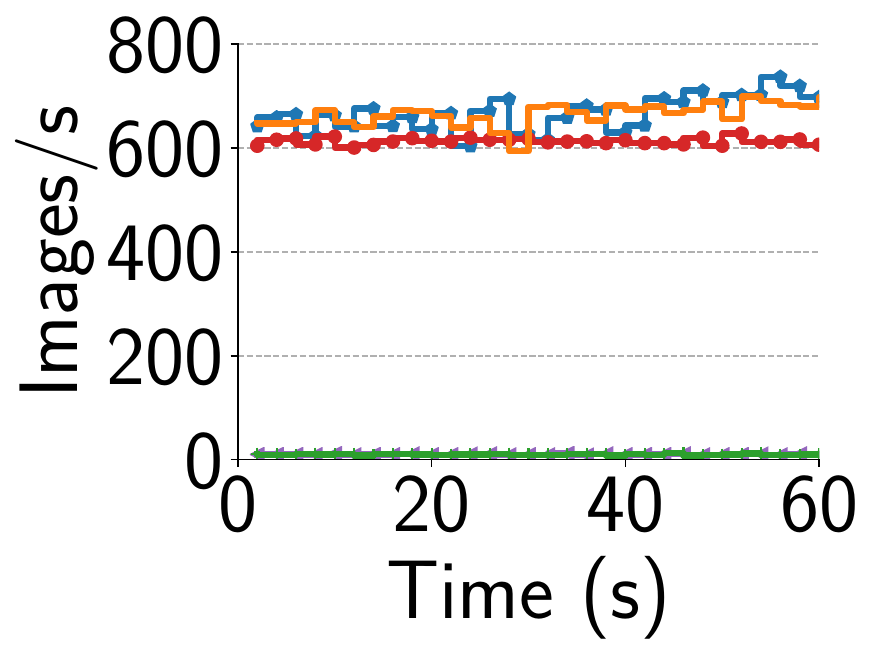}
        \vspace{-6mm}
        \caption{Steady, high load}
        \label{fig:exp-autoscaling-b}
    \end{subfigure}
    \hspace{-1.5mm}%
    \begin{subfigure}[t]{.33\linewidth}
        \centering
        \includegraphics[width=1.0\linewidth]{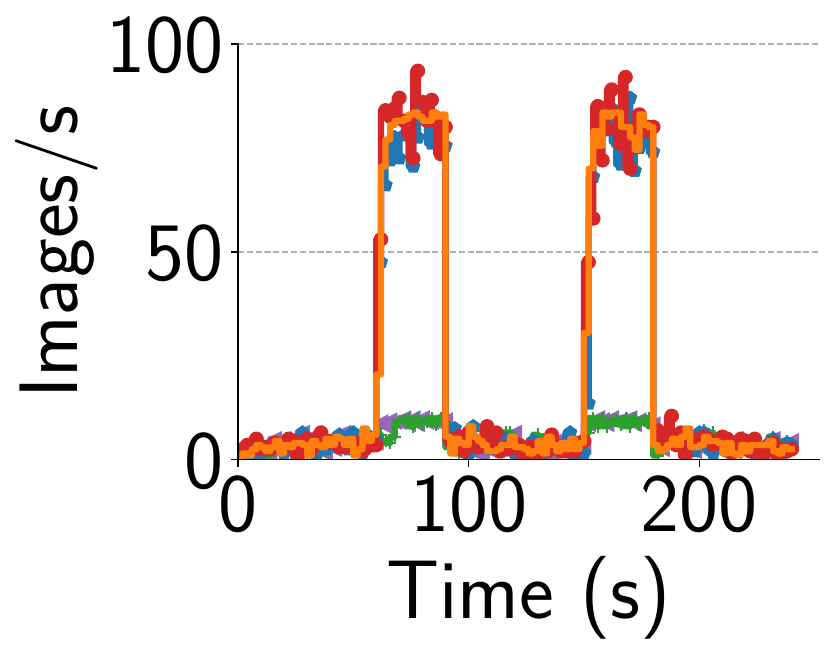}
        \vspace{-6mm}
        \caption{Fluctuating load}
        \label{fig:exp-autoscaling-c}
    \end{subfigure}
    \\
    \begin{subfigure}[t]{0.99\linewidth}
        \centering
        \includegraphics[width=0.93\linewidth]{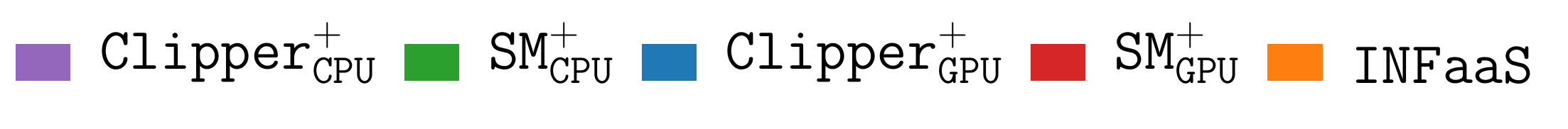}
    \end{subfigure}
    \vspace{-1mm}
    \\
    \begin{subfigure}[t]{.315\linewidth}
        \centering
        \includegraphics[width=1.0\linewidth]{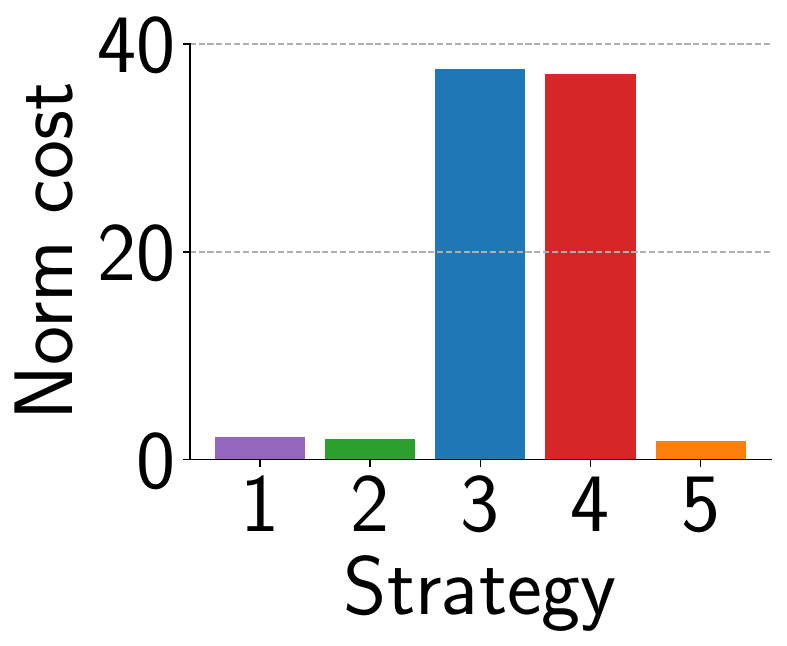}
        \vspace{-6mm}
        \caption{Flat, low load}
        \label{fig:exp-autoscaling-d}
    \end{subfigure}
    \hfill
    \begin{subfigure}[t]{.332\linewidth}
        \centering
        \includegraphics[width=1.0\linewidth]{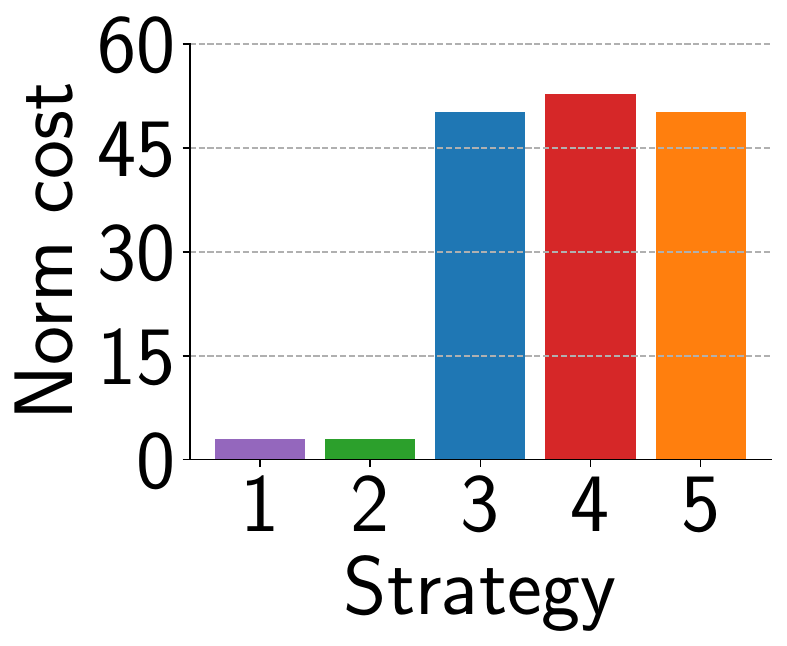}
        \vspace{-6mm}
        \caption{Steady, high load}
        \label{fig:exp-autoscaling-e}
    \end{subfigure}
    \hfill
    \begin{subfigure}[t]{.332\linewidth}
        \centering
        \includegraphics[width=1.0\linewidth]{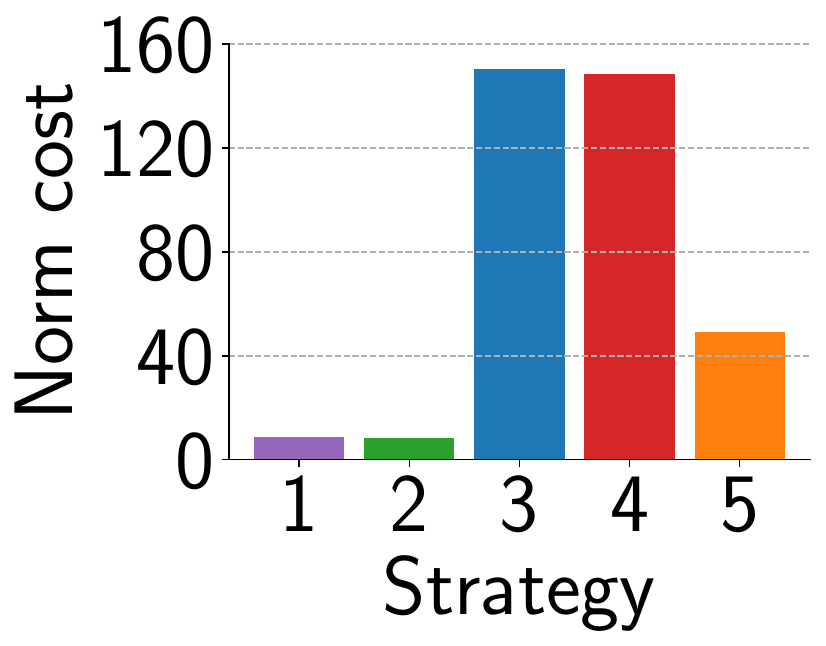}
        \vspace{-6mm}
        \caption{Fluctuating load}
        \label{fig:exp-autoscaling-f}
    \end{subfigure}
    \caption{
     Throughput (top) and cost (bottom), with ResNet50 and batch-1 requests.
     \infaas reduced cost and met the load requirement. %
    }
    \label{fig:exp-autoscaling}
\end{figure}
 
Next, we show the efficiency of \infaas' \modvar selection policy to select and vertically scale the variants. %

\noindent\textbf{Experimental setup. }
To compare \infaas with common configurations developers would choose today, we considered two cases for a model: only GPU variants are used (\staticgpu, \indivgpu) and only CPU variants are used (\staticcpu, \indivcpu).
We used variants derived from one model architecture, ResNet50, and one worker.
\staticcpu pre-loads and persists 2 instances of the TensorFlow CPU variant.
\staticgpu persists one TensorRT variant optimized for batch size of 8, configured to serve the provided peak load by adaptive batching.
\indivcpu horizontally scales the CPU variant. 
\indivgpu horizontally scales a batch-1 optimized TensorRT variant (the cheapest GPU variant).
We measured throughput every 2 seconds, and calculated the total cost. 
The cost for a running variant instance is estimated based on AWS EC2 pricing~\cite{EC2_price}, proportional to its memory footprint.
We normalize cost to 0.031 per GB/s for CPU, 0.190 per GB/s for Inferentia, and 0.498 per GB/s for GPU.

\noindent\textbf{Workloads.} 
We used three query patterns that are commonly observed in real-world setups~\cite{Quasar}: (a) a flat, low load (4 QPS), (b) a steady, high load (slowly increase from 650 to 700 QPS), and (c) a fluctuating load (ranging between 4 and 80 QPS).

\noindent\textbf{Results and discussion.} 
Figures~\ref{fig:exp-autoscaling-a} and~\ref{fig:exp-autoscaling-d} show the throughput and total cost, respectively, for \infaas and the baselines when serving a flat, low load. 
While all systems met this low throughput demand, \staticgpu and \indivgpu incurred high costs since they use GPUs.
\infaas automatically selected CPU variants as they met the demand, thus reducing cost by 21.6$\times$ and 21.3$\times$ compared to \staticgpu and \indivgpu, respectively. 
For a steady, high load (Figures~\ref{fig:exp-autoscaling-b} and~\ref{fig:exp-autoscaling-e}), the observed throughput of \staticcpu and \indivcpu (about 10 QPS) was significantly lower than the demand. 
\infaas automatically selected the batch-8 GPU variant, and both \infaas and \staticgpu met the throughput demand.
While \indivgpu replicated to 2 batch-1 GPU variants to meet the load, it was 5\% more expensive than \infaas and served 15\% fewer QPS.
Finally, for a fluctuating load (Figures~\ref{fig:exp-autoscaling-c} and~\ref{fig:exp-autoscaling-f}), \infaas, \staticgpu, and \indivgpu met the throughput demand, while both \indivcpu and \staticcpu served only 10 QPS.
During low load periods, \infaas selected a CPU variant.
At load spikes (60-90 and 150-180 seconds), \infaas upgraded to an Inferentia batch-1 variant. 
Hence, on average, \infaas was 3$\times$ cheaper than \indivgpu and \staticgpu.
We found that if \infaas were limited to CPU and GPU variants, it would still save 1.7$\times$ cost over both the baselines.
Similarly, even if we allowed baselines to use Inferentia, \infaas would still save 1.9$\times$ cost because baselines cannot dynamically switch between Inferentia and CPU variants. %
\emph{Thus, leveraging variants optimized for different hardware through model-vertical scaling, \infaas is able to adapt to changes in load and query patterns, and improve cost by up to 21.6$\times$ (10$\times$ on average). } 

\subsection{Effectiveness of sharing resources} \label{sec:shareresources} 
\vspace{-1mm}

\begin{figure}[t]
\centering
    \begin{subfigure}[t]{0.5\linewidth}
        \centering
        \includegraphics[width=1.0\linewidth]{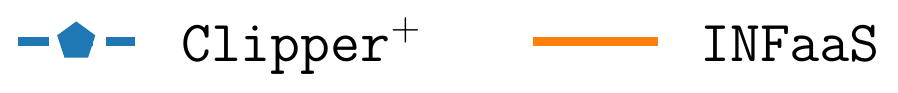}
    \end{subfigure}
    \\ \vspace{-1.5mm}
    \begin{subfigure}[t]{0.49\linewidth}
        \centering
        \includegraphics[width=1.0\linewidth]{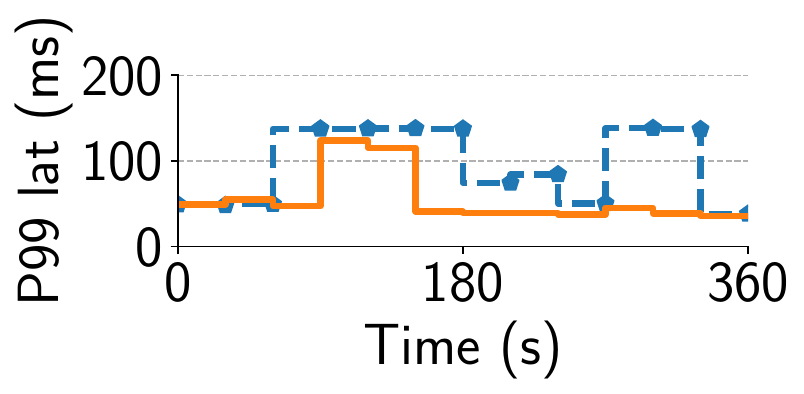}
        \vspace{-7mm}
        \caption{Inception-ResNetV2}
        \label{fig:online_sharing_inception_lat}
    \end{subfigure}
    \hfill
    \begin{subfigure}[t]{0.49\linewidth}
        \centering
        \includegraphics[width=1.0\linewidth]{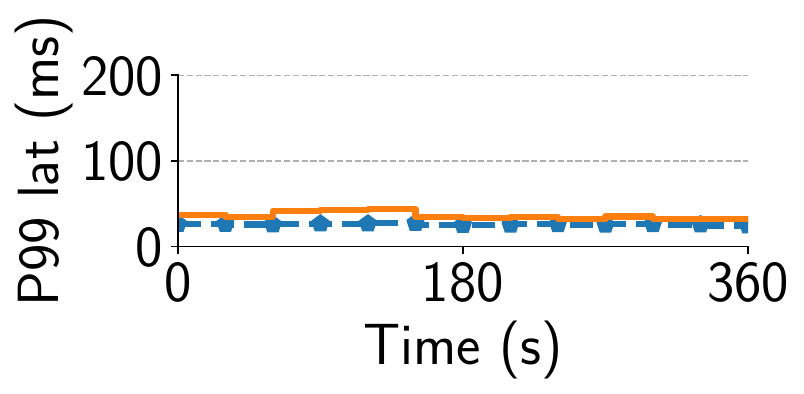}
        \vspace{-7mm}
        \caption{ MobileNetV1}
        \label{fig:online_sharing_mobilenet_lat}
    \end{subfigure}
    \\
    \begin{subfigure}[t]{0.49\linewidth}
        \centering
        \includegraphics[width=1.0\linewidth]{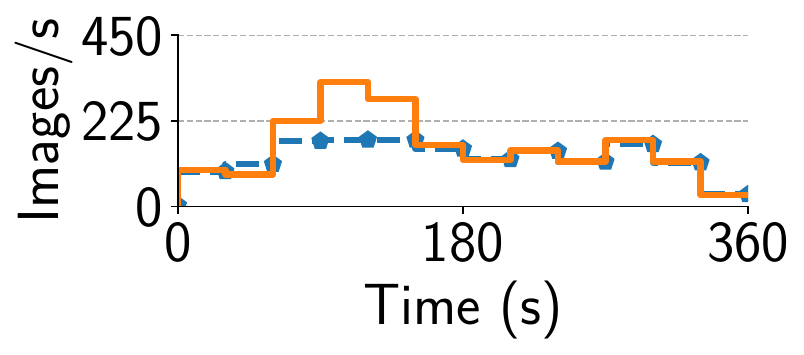}
        \vspace{-7mm}
        \caption{ Inception-ResNetV2}
        \label{fig:online_sharing_inception_qps}
    \end{subfigure}
    \hfill
        \begin{subfigure}[t]{0.49\linewidth}
        \centering
        \includegraphics[width=1.0\linewidth]{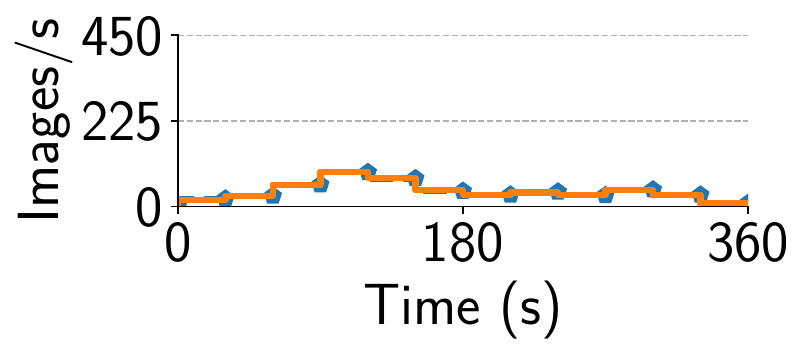}
        \vspace{-7mm}
        \caption{MobileNetV1}
        \label{fig:online_sharing_mobilenet_qps}
    \end{subfigure}
    \caption{
     Performance of co-locating GPU \modvars when 80\% of queries are served by Inception-ResNetV2.  
    }
    \label{fig:online_sharing_incbig}
\end{figure}

\subsubsection{Sharing accelerators}
\vspace{-1mm} 
We now show how \infaas manages and shares accelerators across models without affecting performance of queries. 
We found that co-locating models on an Inferentia chip did not cause noticeable interference, since variants can run on separate cores on the chip.
Thus, we focus on evaluating GPU sharing. 
\infaas detects when \modvars enter the \emph{Overloaded}/\emph{Interfered} state, and either migrates the model to a different GPU, or scales to a new GPU worker if all existing variants on the GPUs are in the \emph{Overloaded}/\emph{Interfered} state. 

\noindent\textbf{Experimental setup. }
We used the baseline of \static with one model persisted on each GPU. 
Since \static requires a pre-defined number of workers, we specified 2 GPU workers.
For fairness, \infaas started from one GPU and was allowed to scale up to 2 GPU workers.
As noted in Section~\ref{sec:multi_tenancy}, the load at which sharing of GPUs starts affecting the performance negatively is different across models. 
We selected two \modvars that diverge in inference latency, throughput, and peak memory: Inception-ResNetV2 (large model) and MobileNetV1 (small model). 
Both variants are TensorRT-optimized for batch-1. 
We report throughput and P99 latency, measured every 30 seconds. 

\noindent\textbf{Workloads.} To show the impact of model popularity on resource sharing, we evaluated a scenario with a popular model serving 80\% QPS, and the other serving 20\% QPS. 
We observed similar results with other popularity distributions or different models. 
We mapped the Twitter trace to a range between 50 and 500 QPS for a total of 75,000 batch-1 queries. 

\noindent\textbf{Results and discussion.} 
Figure~\ref{fig:online_sharing_incbig} shows P99 latency and throughput for both models when Inception-ResNetV2 is popular.
When Inception-ResNetV2 and MobileNetV1 exceeded their profiled latencies, \infaas marked them as interfered around 30 and 50 seconds, respectively.
\infaas started a new GPU worker ($\sim$30 seconds start-up latency), created an instance of each model on it, and spread the load for both models across the GPUs. 
The allocated resources for Inception-ResNetV2 with \static were insufficient and led to a significant latency increase and throughput degradation. 
Unlike \static, \infaas could further mitigate the latency increase by adding more GPU workers (limited to two in this experiment). %
Moreover, \infaas saved 10\% cost compared to \static by (a) bin-packing requests across models to one GPU at low load, and (b) only adding GPUs when contentions were detected.

\vspace{-1mm}
\subsubsection{Co-locating online and offline jobs}
\vspace{-2mm}

In this section, we show using spare resources from online queries for offline jobs allows \infaas to improve utilization. 
To maintain performance for online queries, \infaas throttles offline queries when utilization for the underlying worker exceeds a threshold (set to 40\%), or the observed latency for online queries exceeds a \modvar's profiled latency.
Lower thresholds would starve offline queries, while higher thresholds would incur severe interference.

\noindent\textbf{Experimental setup. } We used one model architecture (ResNet50), one CPU worker, and pre-loaded 2 TensorFlow ResNet50 instances on CPU.
Online requests had a 500ms latency SLO, and load fluctuated between 3 to 8 QPS.
One offline request to ResNet50 was submitted at the beginning of the experiment, containing 500 input images. %

\begin{figure}[t]
\centering
    \begin{subfigure}[t]{0.49\linewidth}
        \centering
        \includegraphics[width=1.0\linewidth]{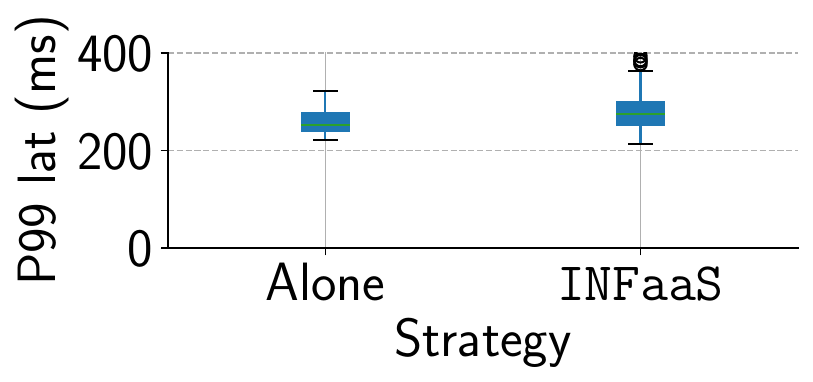}
         \vspace{-7mm}
        \caption{Tail latency for online}
        \label{fig:exp-colocation-a}
    \end{subfigure}
    \hfill
    \begin{subfigure}[t]{0.49\linewidth}
        \centering
        \includegraphics[width=1.0\linewidth]{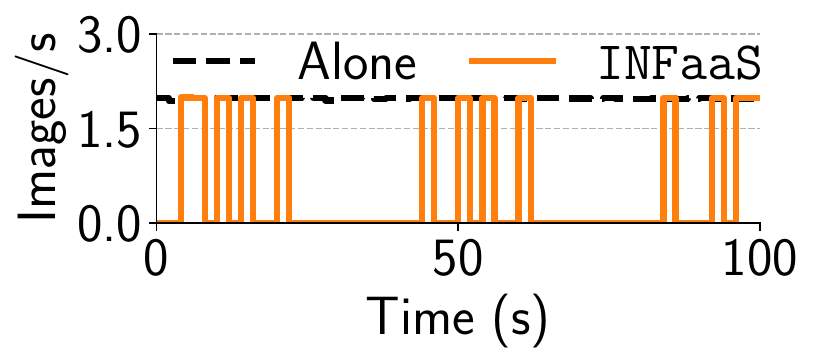}
        \vspace{-7mm}
        \caption{Throughput for offline}
        \label{fig:exp-colocation-b}
    \end{subfigure}
    \\
    \begin{subfigure}[t]{0.49\linewidth}
        \centering
        \includegraphics[width=1.0\linewidth]{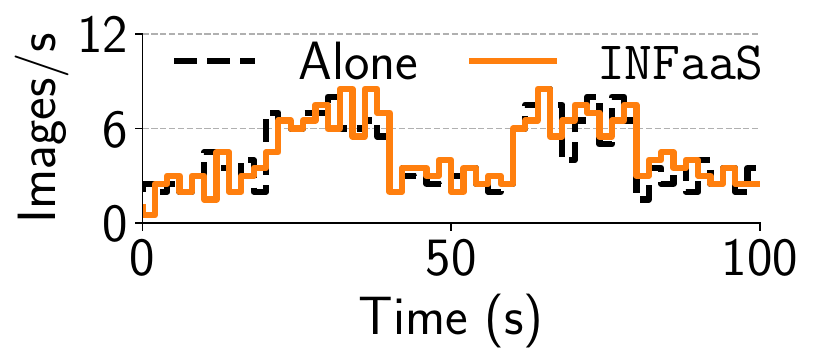}
        \vspace{-7mm}
        \caption{Throughput for online}
        \label{fig:exp-colocation-c}
    \end{subfigure}
    \hfill
    \begin{subfigure}[t]{0.49\linewidth}
        \centering
        \includegraphics[width=1.0\linewidth]{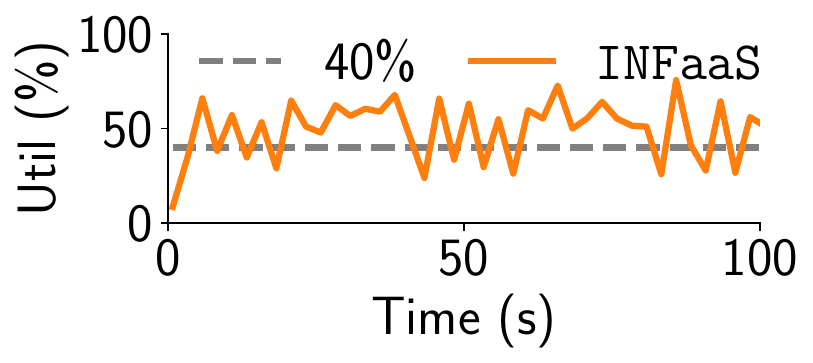}
        \vspace{-7mm}
        \caption{Worker CPU utilization}
        \label{fig:exp-colocation-d}
    \end{subfigure}
    \caption{Performance and utilization of online-offline queries with ResNet50. 
     \emph{Alone}: Serving either online or offline queries, but not both; \infaas: Serving both. 
    }
    \label{fig:exp-colocation}
\end{figure}

\noindent\textbf{Results and discussion.} 
Figures~\ref{fig:exp-colocation-a} to~\ref{fig:exp-colocation-c} contrast the performance of online and offline queries when running alone vs co-located by \infaas. 
Figure~\ref{fig:exp-colocation-d} shows the CPU utilization change for \infaas; the 40\% threshold is marked.
\infaas maintained performance for online requests in both cases by dynamically throttling offline queries. 
On average, 52 images were processed by offline queries during the co-locating period (0-100 seconds).
\infaas throttled offline queries for two long periods (Figure~\ref{fig:exp-colocation-b}): 20-40 and 60-80 seconds, due to high online resource utilization (60\% -- 70\%). 
\subsection{\binfaas' decision overhead} \label{sec:exp-decisionmaking}
\vspace{-1mm}

On the critical path of serving a query, \infaas makes the following decisions: (a) selecting a \modvar and (b) selecting a worker. 
Table~\ref{tab:dec_overhead} shows the median latency of making these decisions for \infaas and the speedup over a brute-force search.
Each row corresponds to a query specifying (1) a registered model and (2) application requirements. 
For each query, we show the decision latency when the selected variant was in (a) \emph{Inactive}, \emph{Overloaded}, or \emph{Interfered} state, and (b) \emph{Active} state. 
These decisions are made using the \modvar selection policy (Section~\ref{subsec:select-modvar}). 

When the registered model was explicitly specified, \infaas incurred low overheads ($\sim$1ms), as it needed to select only a worker.
When the application requirements were provided, and the selected variant was not already loaded (State (a)), \infaas selected a variant and the least-loaded worker for serving in 3.5ms. 
Otherwise, when the selected variant was already loaded (State (b)), \infaas' decision latency for selecting the variant and a worker was 2.2ms.
\emph{\infaas maintains low overheads across its different query submission modes: up to 44$\times$ (35.5$\times$ on average) faster than a brute-force search.}

\begin{table}[t]
  \centering
  \footnotesize
  \resizebox{\linewidth}{!}{
  \begin{tabular}{@{}p{2.2cm} p{2cm} p{1.4cm} p{1.3cm}@{}}\hline
  \textbf{Query}& \textbf{Variant Picked \newline (Valid Options)} & \multicolumn{2}{>{\centering}p{3cm}}{\textbf{Median Latency in ms\newline(Speedup vs brute-force)}} \\ \hline
  & & \textbf{State (a)} & \textbf{State (b)} \\ \hline
  resnet50-trt & resnet50-trt (1) & 1.0 (N/A) & 0.9 (N/A) \\ \hline
  appID, >72\%, 20ms & inceptionv3-trt (5) & 3.5 (27$\times$) & 2.2 (44$\times$) \\ \hline
  \end{tabular}
  } %
  \caption{Median latency and speedup of making variant and worker selection decisions across 3 runs.
  State (a): variants are in the \emph{Inactive} state, \emph{Overloaded} state, or \emph{Interfered} state.
  State (b): variants are in the \emph{Active} state.
  }
  \label{tab:dec_overhead}
\end{table}

\section{Related Work} \label{sec:related-work}
\vspace{-1mm}

\noindent{\bf{Inference serving systems. }} 
TensorFlow Serving~\cite{TensorFlowServing} provided one of the first production environments for models trained using the TensorFlow framework.
Clipper~\cite{Clipper} generalized it to enable the use of different frameworks and application-level SLOs.
Pretzel~\cite{pretzel} built upon Clipper for optimizing the pipelines of inference serving. 
SageMaker~\cite{SageMaker}, AI Platform~\cite{CloudML}, and Azure ML~\cite{Azure_ML} offer developers separate online and offline services that autoscale VMs based on load. 
Triton Inference Server~\cite{TRT_Server} optimizes GPU inference serving, supports CPU models, but requires static model instance configuration.
DeepRecSys~\cite{gupta2020deeprecsys} statically optimizes batching and hardware selection for recommender systems, but  requires developers to specify a variant, manage and scale model resources as the load varies. %
Tolerance Tiers~\cite{OneSizeAll} allows developers to programmatically trade accuracy off for latency. 
None of these existing systems offer a simple model-less interface, like \infaas, to navigate the variant search space on developers' behalf, or dynamically leverage \modvars to meet applications' diverse requirements.

\noindent{\bf{\Modvar generators. }}
Model graph optimizers~\cite{NEO,tensorRT,TVM,tf-xla} perform optimizations, such as quantization and layer fusion, to improve latency and resource usage.
However, developers still need to manually create and select variants, and manage the deployed variants.
\infaas uses these optimizers to create variants that can be used for meeting diverse application requirements, and automates \modvar selection for each query to minimize cost as load and resources vary.

\noindent{\bf{Scaling. }}
Autoscale~\cite{Autoscale} reviewed scaling techniques and argued for a simple approach that maintains the right amount of slack resources while meeting SLOs. %
Similarly, \infaas' autoscalers maintain headrooms and scale-down counters to cautiously scale resources.
MArk~\cite{mark} proposed SLO-aware model scheduling and scaling by using AWS Lambda to absorb unpredictable load bursts.
Existing systems~\cite{gujarati2017swayam,CloudML,SageMaker,Azure_ML} only support VM-level and model-horizontal scaling, while \infaas introduces model-vertical scaling that leverages multiple diverse variants.

\noindent{\bf{Sharing accelerators. }}
NVIDIA MPS~\cite{MPS} enabled efficient sharing of GPUs that facilitated initial exploration into sharing GPUs for deep-learning.
Existing systems~\cite{TRT_Server,TRIMS,Salus,space_time_gpu} also explored how to share GPUs spatially, temporally, or both.
Future NVIDIA GPUs will support MIG~\cite{MIG}: hardware partitions and full isolation.
AWS Inferentia supports spatial and temporal sharing via Neuron SDK~\cite{aws_neuron}.
\infaas' current implementation builds on Triton Inference Server (GPUs) and Neuron SDK (Inferentia), and provides SLO-aware accelerator sharing.
\infaas can also be extended to leverage other mechanisms for sharing additional hardware resources.

\vspace{-1mm} 
\section{Conclusion and Future Work} 
\vspace{-2mm} 

We presented \infaas: a model-less and managed system for distributed inference serving. 
\infaas' model-less interface allows application developers to specify high-level performance, cost or accuracy requirements for queries, leaving \infaas to select and deploy the \modvar, hardware, and scaling configuration. 
\infaas automatically provisions and manages resources for serving inference queries to meet their high-level goals. 
We demonstrated that \infaas' \modvar selection policy and resource sharing leads to reduced costs, better throughput, and fewer SLO violations compared to state-of-the-art inference serving systems.

Going forward, we plan to explore the challenges of managing reconfigurable hardware architectures, such as FPGAs and CGRAs~\cite{virtual_fpga,amorphos,prabhakar2017plasticine}, in a model-less setting, and how they can be utilized to meet diverse application requirements.
Secure and privacy-preserving ML is in its nascent stage~\cite{shredder,liu2017oblivious,juvekar2018gazelle}; we plan to investigate the implications of privacy requirements from applications on model-less inference serving. 

\bibliographystyle{plain}
{\footnotesize
\bibliography{sigplan}}

\begin{thebibliography}{10}

\bibitem{Azure_ML}
{\em Azure Machine Learning}, 2018.
\newblock \url{https://docs.microsoft.com/en-us/azure/machine-learning/}.

\bibitem{grpc}
{\em gRPC}, 2018.
\newblock \url{https://grpc.io/}.

\bibitem{tensorRT}
{\em NVIDIA TensorRT: Programmable Inference Accelerator}, 2018.
\newblock \url{https://developer.nvidia.com/tensorrt}.

\bibitem{Redox}
{\em Redox}, 2018.
\newblock \url{https://github.com/hmartiro/redox}.

\bibitem{TensorFlowServing}
{\em TensorFlow Serving for model deployment in production}, 2018.
\newblock \url{https://www.tensorflow.org/serving/}.

\bibitem{TRT_Server}
{\em NVIDIA Triton Inference Server}, 2020.
\newblock \url{https://github.com/triton-inference-server/server}.

\bibitem{alibaba-npu}
{Hanguang-800 NPU}.
\newblock \url{https://www.t-head.cn/product/npu}.

\bibitem{cherrypick}
Omid Alipourfard, Hongqiang~Harry Liu, Jianshu Chen, Shivaram Venkataraman,
  Minlan Yu, and Ming Zhang.
\newblock Cherrypick: Adaptively unearthing the best cloud configurations for
  big data analytics.
\newblock In {\em 14th {USENIX} Symposium on Networked Systems Design and
  Implementation ({NSDI} 17)}, pages 469--482, Boston, MA, March 2017. {USENIX}
  Association.

\bibitem{AWSEC2}
{Amazon EC2}.
\newblock \url{https://aws.amazon.com/ec2/}, 2018.

\bibitem{AWSS3}
{Amazon S3}.
\newblock \url{https://aws.amazon.com/s3/}, 2018.

\bibitem{SageMaker}
{Amazon SageMaker}.
\newblock \url{https://aws.amazon.com/sagemaker/}, 2018.

\bibitem{NEO}
{Amazon SageMaker Neo}.
\newblock \url{https://aws.amazon.com/sagemaker/neo/}, 2018.

\bibitem{twitter}
{Twitter Streaming Traces}.
\newblock \url{https://archive.org/details/archiveteam-twitter-stream-2018-04},
  2018.

\bibitem{nlp_1}
Mohammed Attia, Younes Samih, Ali Elkahky, and Laura Kallmeyer.
\newblock Multilingual multi-class sentiment classification using convolutional
  neural networks.
\newblock pages 635--640, Miyazaki, Japan, 2018.

\bibitem{aws_neuron}
{AWS Neuron}.
\newblock \url{https://github.com/aws/aws-neuron-sdk}.

\bibitem{inf_cost}
{Deliver high performance ML inference with AWS Inferentia}.
\newblock
  \url{https://d1.awsstatic.com/events/reinvent/2019/REPEAT_1_Deliver_high_performance_ML_inference_with_AWS_Inferentia_CMP324-R1.pdf}.

\bibitem{EC2_price}
{AWS EC2 Pricing}.
\newblock \url{https://aws.amazon.com/ec2/pricing/on-demand/}, 2018.

\bibitem{Inferentia}
{AWS Inferentia}.
\newblock \url{https://aws.amazon.com/machine-learning/inferentia/}, 2018.

\bibitem{bianco2018benchmark}
Simone Bianco, Remi Cadene, Luigi Celona, and Paolo Napoletano.
\newblock Benchmark analysis of representative deep neural network
  architectures.
\newblock {\em IEEE Access}, 6:64270--64277, 2018.

\bibitem{burns2016borg}
Brendan Burns, Brian Grant, David Oppenheimer, Eric Brewer, and John Wilkes.
\newblock {Borg, omega, and kubernetes}.
\newblock {\em Queue}, 14(1):10, 2016.

\bibitem{ConfluxDB}
Prima Chairunnanda, Khuzaima Daudjee, and M.~Tamer {\"O}zsu.
\newblock Confluxdb: Multi-master replication for partitioned snapshot
  isolation databases.
\newblock {\em PVLDB}, 7:947--958, 2014.

\bibitem{TVM}
Tianqi Chen, Thierry Moreau, Ziheng Jiang, Lianmin Zheng, Eddie Yan, Haichen
  Shen, Meghan Cowan, Leyuan Wang, Yuwei Hu, Luis Ceze, Carlos Guestrin, and
  Arvind Krishnamurthy.
\newblock {TVM}: An automated end-to-end optimizing compiler for deep learning.
\newblock In {\em 13th {USENIX} Symposium on Operating Systems Design and
  Implementation ({OSDI} 18)}, pages 578--594, Carlsbad, CA, 2018. {USENIX}
  Association.

\bibitem{clipper_github}
{Clipper}.
\newblock \url{https://github.com/ucbrise/clipper}.

\bibitem{Clipper}
Daniel Crankshaw, Xin Wang, Giulio Zhou, Michael~J. Franklin, Joseph~E.
  Gonzalez, and Ion Stoica.
\newblock Clipper: {A} low-latency online prediction serving system.
\newblock In {\em 14th {USENIX} Symposium on Networked Systems Design and
  Implementation, {NSDI} 2017, Boston, MA, USA, March 27-29, 2017}, pages
  613--627, 2017.

\bibitem{hydra}
Carlo Curino, Subru Krishnan, Konstantinos Karanasos, Sriram Rao, Giovanni~M.
  Fumarola, Botong Huang, Kishore Chaliparambil, Arun Suresh, Young Chen, Solom
  Heddaya, Roni Burd, Sarvesh Sakalanaga, Chris Douglas, Bill Ramsey, and Raghu
  Ramakrishnan.
\newblock Hydra: a federated resource manager for data-center scale analytics.
\newblock In {\em 16th {USENIX} Symposium on Networked Systems Design and
  Implementation ({NSDI} 19)}, pages 177--192, Boston, MA, February 2019.
  {USENIX} Association.

\bibitem{TRIMS}
Abdul Dakkak, Cheng Li, Simon Garcia~De Gonzalo, Jinjun Xiong, and Wen{-}Mei~W.
  Hwu.
\newblock Trims: Transparent and isolated model sharing for low latency deep
  learning inference in function as a service environments.
\newblock {\em CoRR}, abs/1811.09732, 2018.

\bibitem{davidson2017automated}
Thomas Davidson, Dana Warmsley, Michael Macy, and Ingmar Weber.
\newblock {Automated Hate Speech Detection and the Problem of Offensive
  Language}.
\newblock In {\em Proceedings of the Eleventh International AAAI Conference on
  Web and Social Media (ICWSM 2017)}, 2017.

\bibitem{Quasar}
Christina Delimitrou and Christos Kozyrakis.
\newblock Quasar: Resource-efficient and qos-aware cluster management.
\newblock In {\em Proceedings of the 19th International Conference on
  Architectural Support for Programming Languages and Operating Systems},
  ASPLOS '14, pages 127--144, New York, NY, USA, 2014. ACM.

\bibitem{Imagenet}
Jia Deng, Wei Dong, Richard Socher, Li~jia Li, Kai Li, and Li~Fei-fei.
\newblock Imagenet: A large-scale hierarchical image database.
\newblock In {\em In CVPR}, 2009.

\bibitem{jockey}
Andrew~D. Ferguson, Peter Bodik, Srikanth Kandula, Eric Boutin, and Rodrigo
  Fonseca.
\newblock Jockey: Guaranteed job latency in data parallel clusters.
\newblock In {\em Proceedings of the 7th ACM European Conference on Computer
  Systems}, EuroSys '12, pages 99--112, New York, NY, USA, 2012. ACM.

\bibitem{Catapult_ISCA}
Jeremy Fowers, Kalin Ovtcharov, Michael Papamichael, Todd Massengill, Ming Liu,
  Daniel Lo, Shlomi Alkalay, Michael Haselman, Logan Adams, Mahdi Ghandi,
  Stephen Heil, Prerak Patel, Adam Sapek, Gabriel Weisz, Lisa Woods, Sitaram
  Lanka, Steven~K. Reinhardt, Adrian~M. Caulfield, Eric~S. Chung, and Doug
  Burger.
\newblock A configurable cloud-scale dnn processor for real-time ai.
\newblock In {\em Proceedings of the 45th Annual International Symposium on
  Computer Architecture}, ISCA '18, pages 1--14, Piscataway, NJ, USA, 2018.
  IEEE Press.

\bibitem{Autoscale}
Anshul Gandhi, Mor Harchol-Balter, Ram Raghunathan, and Michael~A Kozuch.
\newblock Autoscale: Dynamic, robust capacity management for multi-tier data
  centers.
\newblock {\em ACM Transactions on Computer Systems (TOCS)}, 30(4):14, 2012.

\bibitem{np-completeness}
Michael~R. Garey and David~S. Johnson.
\newblock {\em Computers and Intractability; A Guide to the Theory of
  NP-Completeness}.
\newblock W. H. Freeman \& Co., USA, 1990.

\bibitem{CloudML}
{Google Cloud AI Platform}.
\newblock \url{https://cloud.google.com/ai-platform/}, 2018.

\bibitem{Tiresias}
Juncheng Gu, Mosharaf Chowdhury, Kang~G. Shin, Yibo Zhu, Myeongjae Jeon, Junjie
  Qian, Hongqiang Liu, and Chuanxiong Guo.
\newblock Tiresias: A {GPU} cluster manager for distributed deep learning.
\newblock In {\em 16th {USENIX} Symposium on Networked Systems Design and
  Implementation ({NSDI} 19)}, pages 485--500, Boston, MA, 2019. {USENIX}
  Association.

\bibitem{gujarati2017swayam}
Arpan Gujarati, Sameh Elnikety, Yuxiong He, Kathryn~S McKinley, and Bj{\"o}rn~B
  Brandenburg.
\newblock Swayam: distributed autoscaling to meet slas of machine learning
  inference services with resource efficiency.
\newblock In {\em Proceedings of the 18th ACM/IFIP/USENIX Middleware
  Conference}, pages 109--120. ACM, 2017.

\bibitem{gupta2020deeprecsys}
Udit Gupta, Samuel Hsia, Vikram Saraph, Xiaodong Wang, Brandon Reagen, Gu-Yeon
  Wei, Hsien-Hsin~S. Lee, David Brooks, and Carole-Jean Wu.
\newblock {DeepRecSys: A System for Optimizing End-To-End At-scale Neural
  Recommendation Inference}, 2020.

\bibitem{arch-rmc}
Udit Gupta, Carole-Jean Wu, Xiaodong Wang, Maxim Naumov, Brandon Reagen, David
  Brooks, Bradford Cottel, Kim Hazelwood, Mark Hempstead, Bill Jia, et~al.
\newblock {The Architectural Implications of Facebook's DNN-Based Personalized
  Recommendation}.
\newblock In {\em 2020 IEEE International Symposium on High Performance
  Computer Architecture (HPCA)}, pages 488--501, Feb 2020.

\bibitem{gurobi}
LLC Gurobi~Optimization.
\newblock {Gurobi Optimizer Reference Manual}, 2020.

\bibitem{OneSizeAll}
M.~{Halpern}, B.~{Boroujerdian}, T.~{Mummert}, E.~{Duesterwald}, and
  V.~{Reddi}.
\newblock One size does not fit all: Quantifying and exposing the
  accuracy-latency trade-off in machine learning cloud service apis via
  tolerance tiers.
\newblock In {\em Proceedings of the 19th International Symposium on
  Performance Analysis of Systems and Software (ISPASS)}, 2019.

\bibitem{AppliedML_FB}
Kim Hazelwood, Sarah Bird, David Brooks, Soumith Chintala, Utku Diril, Dmytro
  Dzhulgakov, Mohamed Fawzy, Bill Jia, Yangqing Jia, Aditya Kalro, James Law,
  Kevin Lee, Jason Lu, Pieter Noordhuis, Misha Smelyanskiy, Liang Xiong, and
  Xiaodong Wang.
\newblock Applied machine learning at facebook: A datacenter infrastructure
  perspective.
\newblock In {\em Proceedings of the 2018 IEEE International Symposium on High
  Performance Computer Architecture (HPCA)}, HPCA '18. IEEE, 2018.

\bibitem{mesos}
Benjamin Hindman, Andy Konwinski, Matei Zaharia, Ali Ghodsi, Anthony~D. Joseph,
  Randy Katz, Scott Shenker, and Ion Stoica.
\newblock Mesos: A platform for fine-grained resource sharing in the data
  center.
\newblock In {\em Proceedings of the 8th USENIX Conference on Networked Systems
  Design and Implementation}, NSDI'11, page 295–308, USA, 2011. USENIX
  Association.

\bibitem{focus}
Kevin Hsieh, Ganesh Ananthanarayanan, Peter Bodik, Shivaram Venkataraman,
  Paramvir Bahl, Matthai Philipose, Phillip~B. Gibbons, and Onur Mutlu.
\newblock Focus: Querying large video datasets with low latency and low cost.
\newblock In {\em 13th {USENIX} Symposium on Operating Systems Design and
  Implementation ({OSDI} 18)}, pages 269--286, Carlsbad, CA, October 2018.
  {USENIX} Association.

\bibitem{space_time_gpu}
Paras Jain, Xiangxi Mo, Ajay Jain, Harikaran Subbaraj, Rehan Durrani, Alexey
  Tumanov, Joseph Gonzalez, and Ion Stoica.
\newblock Dynamic space-time scheduling for gpu inference.
\newblock In {\em LearningSys Workshop at Neural Information Processing Systems
  2018}, 2018.

\bibitem{chameleon}
Junchen Jiang, Ganesh Ananthanarayanan, Peter Bodik, Siddhartha Sen, and Ion
  Stoica.
\newblock Chameleon: Scalable adaptation of video analytics.
\newblock In {\em Proceedings of the 2018 Conference of the ACM Special
  Interest Group on Data Communication}, SIGCOMM '18, pages 253--266, New York,
  NY, USA, 2018. ACM.

\bibitem{jing2015visual}
Yushi Jing, David Liu, Dmitry Kislyuk, Andrew Zhai, Jiajing Xu, Jeff Donahue,
  and Sarah Tavel.
\newblock Visual search at pinterest.
\newblock In {\em Proceedings of the 21th ACM SIGKDD International Conference
  on Knowledge Discovery and Data Mining}, pages 1889--1898. ACM, 2015.

\bibitem{pywren}
Eric Jonas, Qifan Pu, Shivaram Venkataraman, Ion Stoica, and Benjamin Recht.
\newblock Occupy the cloud: Distributed computing for the 99\%.
\newblock In {\em Proceedings of the 2017 Symposium on Cloud Computing}, SoCC
  '17, pages 445--451, New York, NY, USA, 2017. ACM.

\bibitem{TPU_ISCA}
Norman~P. Jouppi, Cliff Young, Nishant Patil, David Patterson, Gaurav Agrawal,
  Raminder Bajwa, Sarah Bates, Suresh Bhatia, Nan Boden, Al~Borchers, Rick
  Boyle, Pierre-luc Cantin, Clifford Chao, Chris Clark, Jeremy Coriell, Mike
  Daley, Matt Dau, Jeffrey Dean, Ben Gelb, Tara~Vazir Ghaemmaghami, Rajendra
  Gottipati, William Gulland, Robert Hagmann, C.~Richard Ho, Doug Hogberg, John
  Hu, Robert Hundt, Dan Hurt, Julian Ibarz, Aaron Jaffey, Alek Jaworski,
  Alexander Kaplan, Harshit Khaitan, Daniel Killebrew, Andy Koch, Naveen Kumar,
  Steve Lacy, James Laudon, James Law, Diemthu Le, Chris Leary, Zhuyuan Liu,
  Kyle Lucke, Alan Lundin, Gordon MacKean, Adriana Maggiore, Maire Mahony,
  Kieran Miller, Rahul Nagarajan, Ravi Narayanaswami, Ray Ni, Kathy Nix, Thomas
  Norrie, Mark Omernick, Narayana Penukonda, Andy Phelps, Jonathan Ross, Matt
  Ross, Amir Salek, Emad Samadiani, Chris Severn, Gregory Sizikov, Matthew
  Snelham, Jed Souter, Dan Steinberg, Andy Swing, Mercedes Tan, Gregory
  Thorson, Bo~Tian, Horia Toma, Erick Tuttle, Vijay Vasudevan, Richard Walter,
  Walter Wang, Eric Wilcox, and Doe~Hyun Yoon.
\newblock In-datacenter performance analysis of a tensor processing unit.
\newblock In {\em Proceedings of the 44th Annual International Symposium on
  Computer Architecture}, ISCA '17, pages 1--12, New York, NY, USA, 2017. ACM.

\bibitem{juvekar2018gazelle}
Chiraag Juvekar, Vinod Vaikuntanathan, and Anantha Chandrakasan.
\newblock {GAZELLE}: A low latency framework for secure neural network
  inference.
\newblock In {\em 27th USENIX Security Symposium (USENIX Security 18)}, pages
  1651--1669, 2018.

\bibitem{NoScope}
Daniel Kang, John Emmons, Firas Abuzaid, Peter Bailis, and Matei Zaharia.
\newblock Noscope: Optimizing neural network queries over video at scale.
\newblock {\em Proc. VLDB Endow.}, 10(11):1586--1597, August 2017.

\bibitem{amorphos}
Ahmed Khawaja, Joshua Landgraf, Rohith Prakash, Michael Wei, Eric Schkufza, and
  Christopher~J. Rossbach.
\newblock Sharing, protection, and compatibility for reconfigurable fabric with
  amorphos.
\newblock In {\em 13th {USENIX} Symposium on Operating Systems Design and
  Implementation ({OSDI} 18)}, pages 107--127, Carlsbad, CA, October 2018.
  {USENIX} Association.

\bibitem{Pocket}
Ana Klimovic, Yawen Wang, Patrick Stuedi, Animesh Trivedi, Jonas Pfefferle, and
  Christos Kozyrakis.
\newblock Pocket: Elastic ephemeral storage for serverless analytics.
\newblock In {\em 13th {USENIX} Symposium on Operating Systems Design and
  Implementation ({OSDI} 18)}, pages 427--444, Carlsbad, CA, 2018. {USENIX}
  Association.

\bibitem{pretzel}
Yunseong Lee, Alberto Scolari, Byung-Gon Chun, Marco~Domenico Santambrogio,
  Markus Weimer, and Matteo Interlandi.
\newblock {PRETZEL}: Opening the black box of machine learning prediction
  serving systems.
\newblock In {\em 13th {USENIX} Symposium on Operating Systems Design and
  Implementation ({OSDI} 18)}, pages 611--626, Carlsbad, CA, 2018. {USENIX}
  Association.

\bibitem{hydra_policy}
R.~Levin, E.~Cohen, W.~Corwin, F.~Pollack, and W.~Wulf.
\newblock Policy/mechanism separation in hydra.
\newblock In {\em Proceedings of the Fifth ACM Symposium on Operating Systems
  Principles}, SOSP '75, page 132–140, New York, NY, USA, 1975. Association
  for Computing Machinery.

\bibitem{li2013energy}
Xin Li, Zhuzhong Qian, Sanglu Lu, and Jie Wu.
\newblock Energy efficient virtual machine placement algorithm with balanced
  and improved resource utilization in a data center.
\newblock {\em Mathematical and Computer Modelling}, 58(5-6):1222--1235, 2013.

\bibitem{liu2017oblivious}
Jian Liu, Mika Juuti, Yao Lu, and Nadarajah Asokan.
\newblock Oblivious neural network predictions via minionn transformations.
\newblock In {\em Proceedings of the 2017 ACM SIGSAC Conference on Computer and
  Communications Security}, pages 619--631, 2017.

\bibitem{heracles}
David Lo, Liqun Cheng, Rama Govindaraju, Parthasarathy Ranganathan, and
  Christos Kozyrakis.
\newblock {Heracles: Improving Resource Efficiency at Scale}.
\newblock In {\em Proceedings of the 42Nd Annual International Symposium on
  Computer Architecture}, ISCA '15, pages 450--462, New York, NY, USA, 2015.
  ACM.

\bibitem{themis}
Kshiteej Mahajan, Arjun Balasubramanian, Arjun Singhvi, Shivaram Venkataraman,
  Aditya Akella, Amar Phanishayee, and Shuchi Chawla.
\newblock Themis: Fair and efficient {GPU} cluster scheduling.
\newblock In {\em 17th {USENIX} Symposium on Networked Systems Design and
  Implementation ({NSDI} 20)}, pages 289--304, Santa Clara, CA, February 2020.
  {USENIX} Association.

\bibitem{shredder}
Fatemehsadat Mireshghallah, Mohammadkazem Taram, Prakash Ramrakhyani, Ali
  Jalali, Dean Tullsen, and Hadi Esmaeilzadeh.
\newblock Shredder: Learning noise distributions to protect inference privacy.
\newblock In {\em Proceedings of the Twenty-Fifth International Conference on
  Architectural Support for Programming Languages and Operating Systems},
  ASPLOS '20, page 3–18, New York, NY, USA, 2020. Association for Computing
  Machinery.

\bibitem{hotos_anonymized}
Author names hidden for double-blind~review purpose.
\newblock Anonymized title.
\newblock In {\em Prior Workshop}.

\bibitem{MPS}
{NVIDIA MPS}.
\newblock
  \url{https://docs.nvidia.com/deploy/pdf/CUDA_Multi_Process_Service_Overview.pdf},
  2018.

\bibitem{MIG}
{NVIDIA Multi-instance GPU}.
\newblock \url{https://www.nvidia.com/en-us/technologies/multi-instance-gpu/},
  2020.

\bibitem{Oh:2018}
Young~H. Oh, Quan Quan, Daeyeon Kim, Seonghak Kim, Jun Heo, Sungjun Jung,
  Jaeyoung Jang, and Jae~W. Lee.
\newblock A portable, automatic data quantizer for deep neural networks.
\newblock In {\em Proceedings of the 27th International Conference on Parallel
  Architectures and Compilation Techniques}, PACT '18, pages 17:1--17:14, New
  York, NY, USA, 2018. ACM.

\bibitem{polyzotis2017data}
Neoklis Polyzotis, Sudip Roy, Steven~Euijong Whang, and Martin Zinkevich.
\newblock Data management challenges in production machine learning.
\newblock In {\em Proceedings of the 2017 ACM International Conference on
  Management of Data}, pages 1723--1726. ACM, 2017.

\bibitem{Scanner}
Alex Poms, Will Crichton, Pat Hanrahan, and Kayvon Fatahalian.
\newblock {Scanner: Efficient video analysis at scale}.
\newblock {\em ACM Transactions on Graphics (TOG)}, 37(4):1--13, 2018.

\bibitem{prabhakar2017plasticine}
Raghu Prabhakar, Yaqi Zhang, David Koeplinger, Matt Feldman, Tian Zhao, Stefan
  Hadjis, Ardavan Pedram, Christos Kozyrakis, and Kunle Olukotun.
\newblock {Plasticine: A Reconfigurable Architecture for Parallel Patterns}.
\newblock In {\em 2017 ACM/IEEE 44th Annual International Symposium on Computer
  Architecture (ISCA)}, pages 389--402. IEEE, 2017.

\bibitem{mlperf}
V.~J. {Reddi}, C.~{Cheng}, D.~{Kanter}, P.~{Mattson}, G.~{Schmuelling},
  C.~{Wu}, B.~{Anderson}, M.~{Breughe}, M.~{Charlebois}, W.~{Chou},
  R.~{Chukka}, C.~{Coleman}, S.~{Davis}, P.~{Deng}, G.~{Diamos}, J.~{Duke},
  D.~{Fick}, J.~S. {Gardner}, I.~{Hubara}, S.~{Idgunji}, T.~B. {Jablin},
  J.~{Jiao}, T.~S. {John}, P.~{Kanwar}, D.~{Lee}, J.~{Liao}, A.~{Lokhmotov},
  F.~{Massa}, P.~{Meng}, P.~{Micikevicius}, C.~{Osborne}, G.~{Pekhimenko},
  A.~T.~R. {Rajan}, D.~{Sequeira}, A.~{Sirasao}, F.~{Sun}, H.~{Tang},
  M.~{Thomson}, F.~{Wei}, E.~{Wu}, L.~{Xu}, K.~{Yamada}, B.~{Yu}, G.~{Yuan},
  A.~{Zhong}, P.~{Zhang}, and Y.~{Zhou}.
\newblock Mlperf inference benchmark.
\newblock In {\em 2020 ACM/IEEE 47th Annual International Symposium on Computer
  Architecture (ISCA)}, pages 446--459, 2020.

\bibitem{redis-www}
Redis.
\newblock \url{https://redis.io}, 2018.

\bibitem{mage}
Francisco Romero and Christina Delimitrou.
\newblock Mage: Online and interference-aware scheduling for multi-scale
  heterogeneous systems.
\newblock In {\em Proceedings of the 27th International Conference on Parallel
  Architectures and Compilation Techniques}, PACT '18, New York, NY, USA, 2018.
  Association for Computing Machinery.

\bibitem{binpacking}
Steven~S. Seiden.
\newblock On the online bin packing problem.
\newblock {\em J. ACM}, 49(5):640--671, September 2002.

\bibitem{shahrad2020serverless}
Mohammad Shahrad, Rodrigo Fonseca, Inigo Goiri, Gohar Irfan, Paul Batum, Jason
  Cooke, Eduardo Laureano, Colby Tresness, Mark Russinovich, and Ricardo
  Bianchini.
\newblock {Serverless in the Wild: Characterizing and Optimizing the Serverless
  Workload at a Large Cloud Provider}.
\newblock In {\em 2020 {USENIX} Annual Technical Conference ({USENIX} {ATC}
  20)}, Boston, MA, USA, July 2020. {USENIX} Association.
\newblock To Appear.

\bibitem{nexus}
Haichen Shen, Lequn Chen, Yuchen Jin, Liangyu Zhao, Bingyu Kong, Matthai
  Philipose, Arvind Krishnamurthy, and Ravi Sundaram.
\newblock Nexus: A gpu cluster engine for accelerating dnn-based video
  analysis.
\newblock In {\em Proceedings of the 27th ACM Symposium on Operating Systems
  Principles}, SOSP ’19, page 322–337, New York, NY, USA, 2019. Association
  for Computing Machinery.

\bibitem{nlp_2}
Leonid Velikovich, Ian Williams, Justin Scheiner, Petar~S. Aleksic, Pedro~J.
  Moreno, and Michael Riley.
\newblock Semantic lattice processing in contextual automatic speech
  recognition for google assistant.
\newblock In {\em Interspeech 2018, 19th Annual Conference of the International
  Speech Communication Association, Hyderabad, India, 2-6 September 2018.},
  pages 2222--2226, 2018.

\bibitem{ernest}
Shivaram Venkataraman, Zongheng Yang, Michael Franklin, Benjamin Recht, and Ion
  Stoica.
\newblock Ernest: Efficient performance prediction for large-scale advanced
  analytics.
\newblock In {\em 13th {USENIX} Symposium on Networked Systems Design and
  Implementation ({NSDI} 16)}, pages 363--378, Santa Clara, CA, 2016. {USENIX}
  Association.

\bibitem{von1978bound}
Joachim von~zur Gathen and Malte Sieveking.
\newblock A bound on solutions of linear integer equalities and inequalities.
\newblock {\em Proceedings of the American Mathematical Society},
  72(1):155--158, 1978.

\bibitem{WMT16}
{ACL 2016 First Conference on Machine Translation (WMT16)}.
\newblock \url{http://www.statmt.org/wmt16/}.

\bibitem{Gandiva}
Wencong Xiao, Romil Bhardwaj, Ramachandran Ramjee, Muthian Sivathanu, Nipun
  Kwatra, Zhenhua Han, Pratyush Patel, Xuan Peng, Hanyu Zhao, Quanlu Zhang, Fan
  Yang, and Lidong Zhou.
\newblock Gandiva: Introspective cluster scheduling for deep learning.
\newblock In {\em 13th {USENIX} Symposium on Operating Systems Design and
  Implementation ({OSDI} 18)}, pages 595--610, Carlsbad, CA, 2018. {USENIX}
  Association.

\bibitem{tf-xla}
{XLA: Optimizing Compiler for Machine Learning}.
\newblock \url{https://www.tensorflow.org/xla}.

\bibitem{Salus}
Peifeng Yu and Mosharaf Chowdhury.
\newblock Salus: Fine-grained {GPU} sharing primitives for deep learning
  applications.
\newblock {\em CoRR}, abs/1902.04610, 2019.

\bibitem{virtual_fpga}
Yue Zha and Jing Li.
\newblock Virtualizing fpgas in the cloud.
\newblock In {\em Proceedings of the Twenty-Fifth International Conference on
  Architectural Support for Programming Languages and Operating Systems},
  ASPLOS ’20, page 845–858, New York, NY, USA, 2020. Association for
  Computing Machinery.

\bibitem{mark}
Chengliang Zhang, Minchen Yu, Wei Wang, and Feng Yan.
\newblock Mark: Exploiting cloud services for cost-effective, slo-aware machine
  learning inference serving.
\newblock In {\em 2019 {USENIX} Annual Technical Conference ({USENIX} {ATC}
  19)}, pages 1049--1062, Renton, WA, July 2019. {USENIX} Association.

\bibitem{videostorm}
Haoyu Zhang, Ganesh Ananthanarayanan, Peter Bodik, Matthai Philipose, Paramvir
  Bahl, and Michael~J. Freedman.
\newblock Live video analytics at scale with approximation and delay-tolerance.
\newblock In {\em 14th {USENIX} Symposium on Networked Systems Design and
  Implementation ({NSDI} 17)}, pages 377--392, Boston, MA, March 2017. {USENIX}
  Association.

\end{thebibliography}

\begin{appendices}
  \clearpage
\section{Model Architectures \& Variants}
\label{sec:detail-modarch-vars}

Table~\ref{tab:mod_arch_details} shows details of the models we used for evaluating \infaas. 
Model architectures within the same model family (corresponding to Table~\ref{tab:mod_arch_vars}) are grouped together.
The number of layers and parameters correspond to the unoptimized, vanilla implementation of the respective model architecture.
Image classification top-1 accuracies are for the ImageNet dataset, and translation BLEU scores (denoted by *) are for the WMT16 English-German dataset.

\begin{table}[ht!]
  \centering
  \begin{tabular}{@{}p{2cm}llll@{}}
\toprule
\textbf{Model Arch.}                                          & \textbf{Acc.} & \textbf{\#Vars} & \textbf{\#Layers} & \textbf{\#Params} \\ \midrule
MobileNetV1                                                   & 70.4          & 10              & 30                & 4,253,864         \\
MobileNetV2                                                   & 71.3          & 3               & 30                & 3,538,984         \\ \midrule
AlexNet                                                       & 56.6          & 9               & 8                 & 62,378,344        \\ \midrule
DenseNet121                                                   & 75.0          & 12              & 121               & 8,062,504         \\
DenseNet169                                                   & 76.2          & 5               & 169               & 14,307,880        \\
DenseNet201                                                   & 77.3          & 5               & 201               & 20,242,984        \\ \midrule
ResNet50                                                      & 74.9          & 19              & 50                & 25,636,712        \\
ResNet50V2                                                    & 76.0          & 4               & 50                & 25,613,800        \\
ResNext50                                                     & 77.7          & 3               & 50                & $\sim$25,700,000  \\
ResNet101                                                     & 76.4          & 12              & 101               & 44,707,176        \\
ResNet101V2                                                   & 77.2          & 4               & 101               & 44,675,560        \\
ResNext101                                                    & 78.7          & 3               & 101               & $\sim$45,300,000  \\
ResNet152                                                     & 76.6          & 12              & 152               & 60,419,944        \\
ResNet152V2                                                   & 78.0          & 4               & 152               & 60,380,648        \\ \midrule
VGG16                                                         & 71.3          & 18              & 16                & 138,357,544       \\
VGG19                                                         & 71.3          & 12              & 19                & 143,667,240       \\ \midrule
InceptionV3                                                   & 77.9          & 12              & 48                & 23,851,784        \\
Xception                                                      & 79.0          & 3               & 76                & 22,910,480        \\
Inception\-/ResNetV2 & 80.3          & 10              & 164               & 55,873,736        \\ \midrule
NasNetMobile                                                  & 74.4          & 3               & 914               & 5,326,716         \\
NasNetLarge                                                   & 82.5          & 3               & 1244              & 88,949,818        \\ \midrule
GNMT                                                          & 24.5*         & 9               & 16                & $\sim$210,000,000 \\ \bottomrule
\end{tabular}
  \caption{\small Models used for evaluating \infaas.}
  \label{tab:mod_arch_details}
\end{table}

\end{appendices}

\end{document}